\documentclass[lettersize,journal]{IEEEtran}
\usepackage{amsmath,amsfonts}
\usepackage{algorithmic}
\usepackage{array}
\usepackage[caption=false,font=normalsize,labelfont=sf,textfont=sf]{subfig}
\usepackage{textcomp}
\usepackage{stfloats}
\usepackage{url}
\usepackage{verbatim}
\usepackage{graphicx}
\def\BibTeX{{\rm B\kern-.05em{\sc i\kern-.025em b}\kern-.08em
    T\kern-.1667em\lower.7ex\hbox{E}\kern-.125emX}}
\usepackage{balance}
\usepackage{xspace}
\newcommand{\sysname}{RFDrone\xspace}
\begin{document}

\title{RFID-based Relative Localization for Nanodrone Swarm Platooning}


\author{Wei Sun \\
\IEEEauthorblockA{Wright State University, Dayton, Ohio, USA}
}



\maketitle

\begin{abstract}
A nanodrone swarm is formulated by multiple lightweight and low-cost nanodrones to perform tasks in very challenging environments. Therefore, it is essential to recognize the relative positions of the nanodrones in the swarm for accurate and safe platooning. However, the camera or LiDAR sensors suffer from the line-of-sight (LoS) sensing. This paper presents the design, implementation, and evaluation of \sysname, a system that can enable the relative positioning for the nanodrone swarm using the commodity passive RFID tags attached to each nanodrone for safe and accurate platooning. To do so, we propose a novel algorithm for spatial-temporal phase profiling based on the swarm platooning nature and further design a machine-learning model to predict the relative position of the tagged nanodrones in the swarm. Our experiments show that \sysname can accurately estimate the relative positions of nanodrones in the swarm, whereby the 3D geometry of the nanodrone swarm pattern is easily obtained across different environments, tag diversity, and swarm pattern diversity. 
\end{abstract}


\section{Introduction}



\begin{figure}
\centering
\captionsetup{width=0.22\textwidth}
\begin{minipage}{0.25\textwidth}
 \centering
\includegraphics[width=0.8\linewidth]{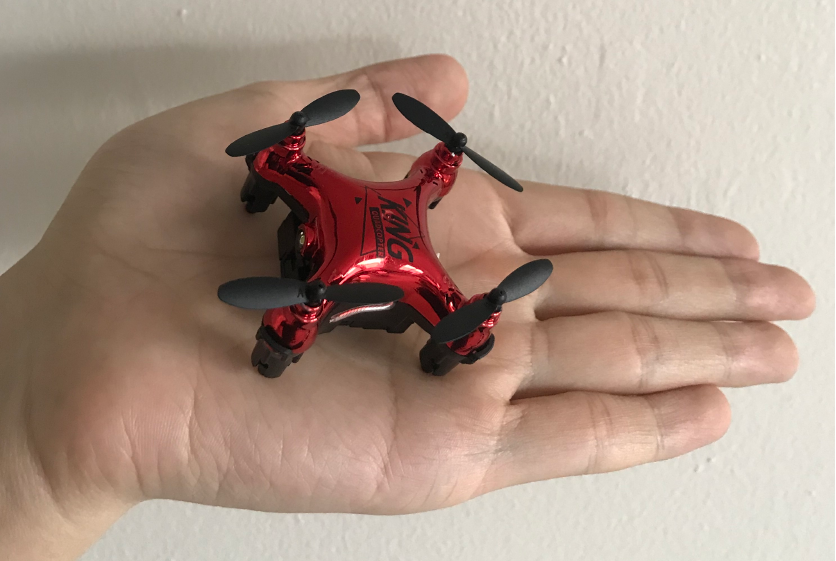}
    \caption{The nanodrone with size of 70x48x35mm, fitting in the palm.}
    \label{fig:nanodrone}
\end{minipage}%
\begin{minipage}{0.25\textwidth}
  \centering
  \includegraphics[width=0.8\linewidth]{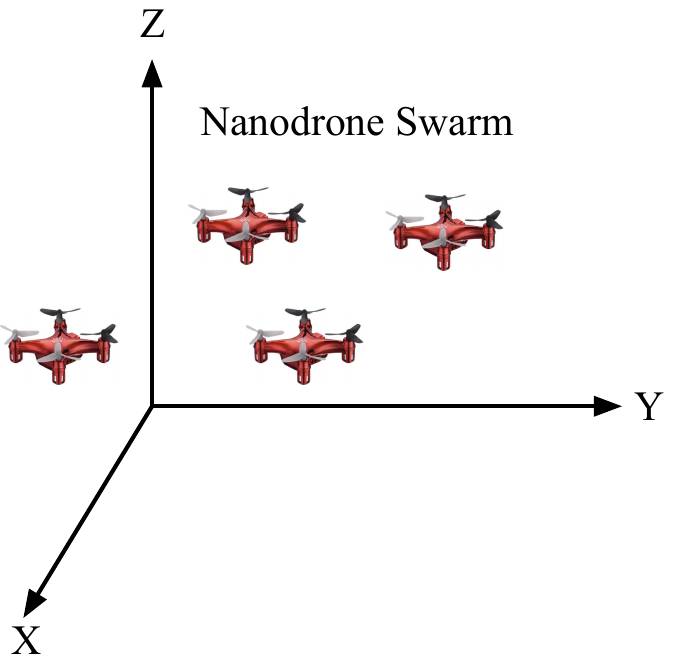}
    \caption{Example of nanodrone swarm with four nanodrones flying in the 3D space.}
    \label{fig:drone:swarm}
\end{minipage} 
\end{figure}

\begin{figure}
  \centering
  \includegraphics[width=0.5\linewidth]{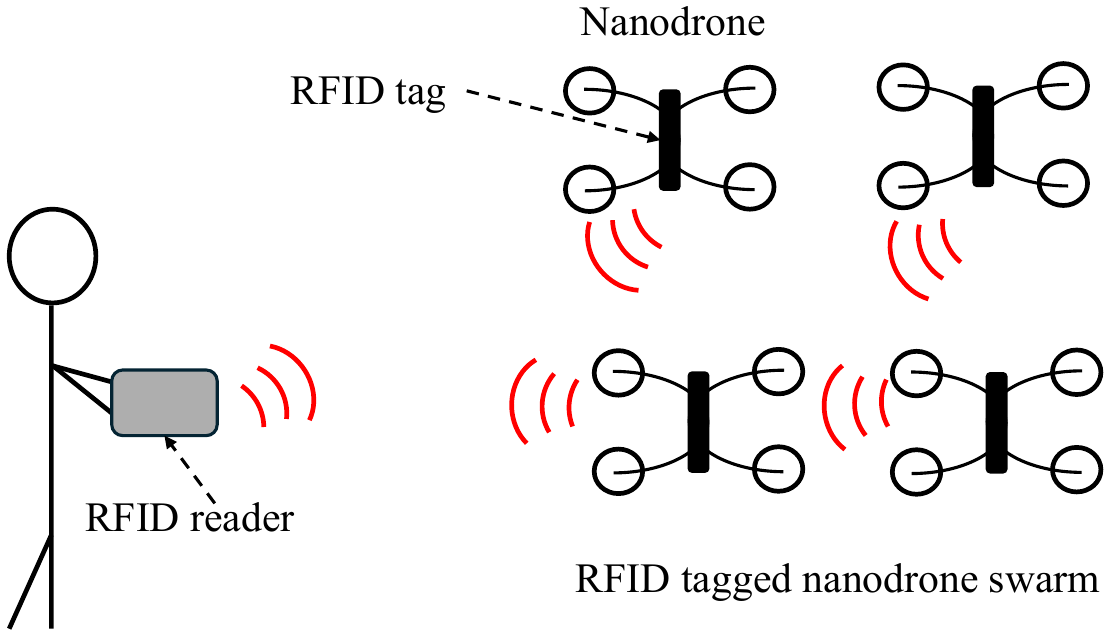}
    \caption{\sysname: Relative positioning nanodrone swarm with commodity passive RFID tags for human-drone swarm interaction.}
    \label{fig:human:drone}
\end{figure}

Recently, the lightweight and small form-factor nanodrone (e.g., quadrotor or quadcopter) is designed, unlike the large industrial drones (e.g., Amazon drones and DJI drones have orders of magnitude larger size), which can just fit in our palm as shown in Fig.~\ref{fig:nanodrone}. The nanodrone is as small as 22x22x20 mm (e.g., Mini Drone FY804~\cite{fy804}) and  70x48x35 mm (e.g., Holyton HT02 Mini Drone~\cite{holy}), which can enable it to traverse through the small, tight and narrow space such as plant canopies and crevices to perform the tasks. So, we can use nanodrones to perform sensing tasks in some severe environments (e.g., forests and disaster sites) to rescue lives~\cite{xiang2016design}. These nanodrones proliferate the entertainment industry~\cite{du2019fast}, exhibiting the exceptional agility and flexibility of the nanodrones shows~\cite{waibel2017drone, drone_show}. We can also enable the human-nanodrone interactions as shown in Fig.~\ref{fig:human:drone}. However, we cannot instrument different sensors on the nanodrone for sensing purposes due to its small size, thereby the ability of nanodrone to perform the tasks is limited. Inspired by the concept of crowdsourcing~\cite{brabham2013crowdsourcing}, multiple nanodrones can formulate a nanodrone swarm for emergency response and hazard detection in urban settings~\cite{loianno2016estimation}. As such, it is important for the nanodrone swarm to formulate an accurate geometry of swarm pattern for safe and accurate platooning~\cite{chen2015safe}, such that the sensing tasks can be efficiently performed. 


To recognize the relative position of nanodrones in the swarm as shown in Fig.~\ref{fig:drone:swarm}, the straightforward idea is to use the computer vision-based sensors ( e.g., stereo cameras) or ranging-based LiDAR sensors~\cite{liu2012fast} for device-free sensing. However, these advanced sensors cannot work properly in non-line-of-sight settings and are not cost-effective~\cite{lu2020milliego}, especially in the presence of airborne obscurants (e.g., dust, fog, and smoke). Alternatively, RF-based sensing techniques have been widely developed in indoor localization, which can be leveraged to recognize the relative positions of the nanodrones. However, it is not possible to instrument these bulky RF radios on the nanodrone. More importantly, the highly cluttered indoor environment poses a great challenge to accurately recognize the relative position of the nanodrones in the swarm for safe and accurate platooning.




In this paper, we propose \sysname, a system that can accurately recognize the relative position of the nanodrones in the swarm with a commodity passive RFID tag attached to each nanodrone in the swarm as shown in Fig.~\ref{fig:human:drone} where the hand-held RFID reader interrogates tagged-nanodrones for human-nanodrone swarm interaction. Our \sysname can accurately localize each nanodrone's position relative to the other nanodrones' positions in the swarm for swarm pattern recognition. Since the commodity passive RFID tags are small form-factor, battery-free, and low cost, they do not affect the nanodrone's platooning and draw energy from the nanodrone. \sysname enables accurate relative localization for nanodrone warm platooning.

However, there are great challenges for relatively positioning nanodrone swarm. Firstly, the nanodrone body is not stable during the flight, especially for the nanodrones without a stabilization system, which can introduce varying artifacts to the backscatter signals. As a result, it is difficult to predict the nanodrone's relative position in the swarm without eliminating these artifacts. Secondly, since there are multiple nanodrones in the swarm platooning in the environment, each of them can reflect off the backscatter signals to enrich the multipath effect, disabling the accurate localization of each nanodrone in the swarm. Lastly, since the RFID reader needs to interrogate all the tags sequentially when the nanodrone swarm is platooning, it is difficult to obtain the complete phase information for all the tags. As a result, it is challenging to predict the nanodrone's relative position based on incomplete phase information.

\begin{figure}
\centering
\captionsetup{width=0.23\textwidth}
\begin{minipage}{0.25\textwidth}
 \centering
  \includegraphics[width=\linewidth]{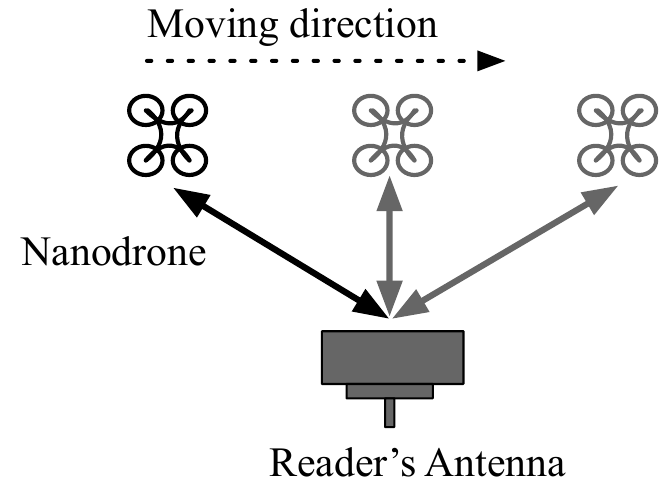}
    \caption{The nanodrone is moving from left to right in front of the reader's antenna.}
    \label{fig:one:drone}
\end{minipage}%
\begin{minipage}{0.25\textwidth}
  \centering
  \includegraphics[width=\linewidth]{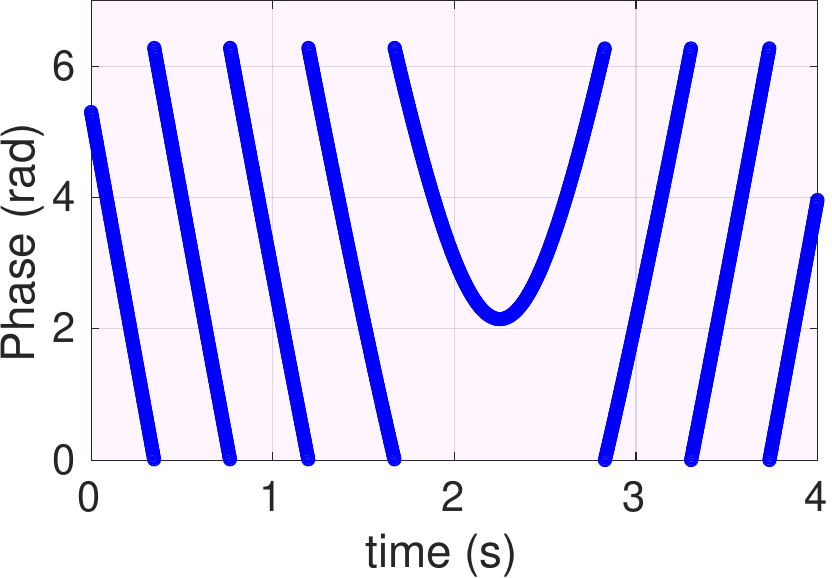}
    \caption{Phase readings over time, when the RFID-tagged nanodrone is moving from left to right in front of the reader's antenna.}
    \label{fig:one:drone:phase}
\end{minipage} 
\end{figure}

To address the above challenges, we propose to profile the tagged nanodrone's spatial-temporal phase information and further leverage machine learning to predict the nanodrone's relative position. Specifically, we harness the nanodrone swarm's platooning principle where each nanodrone in the swarm performs the same platooning operation (e.g., all the nanodrones fly in a specific direction by nature). As such, we can leverage the linear relationship between phase readings and the reader-tag distance to obtain the spatial-temporal phase profile as shown in Fig.~\ref{fig:one:drone} and Fig.~\ref{fig:one:drone:phase}, which can indicate the relative positions of the nanodrones in the swarm across the x, y, and z-axis with the RFID reader in the original coordinates. By platooning nature, the nanodrone swarm should fly along one of three axes, which requires us to match the phase readings with the specific axis. Therefore, we propose to use a camera colocated with the RFID reader for platooning direction estimation. Even though the camera cannot recognize all the nanodrones' flight directions due to the nanodrone-to-nanodrone obstruction, we can easily identify the nanodrone swarm's platooning direction as long as one nanodrone is exposed to the camera. 

To eliminate the effect of the nandorone's body variation, we propose to filter out the phase information affected by the nanodrone's body variation based on the statistical phase readings. This is because all the nanodrones perform the same platooning operation by platooning nature. Then, we interpolate the incomplete phase readings based on the fact that each nanodrone should provide the same pattern of the spatial-temporal phase profile, which can offer us a reliable spatial-temporal phase profile. Furthermore, we design a machine learning model to predict the relative position of the nanodrones in the swarm based on the spatial-temporal phase profile, using bidirectional LSTM that can model the time-related phase readings well.

\noindent \textbf{Contributions.} We prototype \sysname with commodity passive RFID tags. We only need to extract the phase information of the backscattered channel to profile the spatial and temporal phase readings. \sysname's contributions are three-fold:
\begin{itemize}
    \item To the best of our knowledge,~\sysname is the first system that can accurately recognize the 3D geometry of the nanodrone swarm pattern with commodity passive RFID tags through relative localization.
    \item Second, we propose to estimate the relative positions of the nanodrones in the swarm from the spatial-temporal phase profiles to obtain the 3D nanodrone swarm pattern using machine learning. 
    \item Finally, we built and implemented the \sysname with commodity passive RFID tags and demonstrated its ability and accuracy to recognize the 3D geometry of the nanodrone swarm pattern.
\end{itemize}


\section{Related work}
\label{sec:work}


\textbf{RFID-Based Localization.} RFID tags are widely used in indoor localization~\cite{wang2013dude, wang2013rf,sun2021taguard}, intelligent vehicles~\cite{srinivasan2025methods,sun2024rfdrive,sun2022feasibility,sun2020allergie},  health sensing~\cite{sun2023short,sun2022human,xu2021embracing,sun2021healthy,sun2021rfitness,xu2020embracing}, parallel decoding~\cite{sun2021towards}, RFID security~\cite{sun2021destructive,sun2020destructive} due to their low cost and small form factor. Early work~\cite{zhou2009rfid} mainly focuses on using the signal strength and received signal phase for indoor localization, which can achieve good performance in line-of-sight scenarios. However, when the multipath exists in a highly cluttered indoor environment, the constructive and destructive signal addition will deteriorate the received signal strength and phase information.

To deal with the multipath effect in the indoor area, past literature leverages space diversity to resolve the multipath. Specifically, multiple tags are emulated as a tag array to derive the angle-of-arrival of the backscattered signal~\cite{sun2021orientation}, which can resolve the multipath. RF-IDraw~\cite{wang2014rf} exploits the tag spacing to accurately track the finger gestures. PinIt~\cite{wang2013dude} compares the multipath profile of the reference tag and the desired tag by emulating the moving antenna of the reader as an array to resolve the multipath for indoor localization. Tagyro~\cite{wei2016gyro} leverages the tag array attached to the objects for orientation estimation through AoA information. Apart from using the signal strength, signal phase, and AoA, people also use time-of-flight (ToF) to do localization. However, the fine-grained ToF measurements require a high bandwidth, which is not available in commodity passive RFID systems with tens of KHz bandwidth. Recently, RFind~\cite{ma2017minding} emulates a bandwidth of 300MHz for RFID-based sensing, which is $10\times$ larger than the ISM band for UHF RFID communication. However, transmitting FCC-compatible signals with lower power at the band outside of the ISM band will significantly reduce the communication range. In comparison to the above techniques, we target the nanodrone swarm pattern recognition by exploiting the relative positions of nanodrones in the swarm based on the spatial and temporal phase profiles.

\noindent \textbf{Drone Related Applications.} The drone has become an important tool for ubiquitous sensing due to its agility and high mobility. Therefore, there are many drone-based applications for large industrial drones and small nanodrones.

We first see many papers exploiting the reliability of the drone system. For example, Bleep~\cite{bannis2020bleep} leverages the acoustics signal to measure the stability of the drone. SafetyNet~\cite{gowda2016tracking} uses four GPS receivers to track the orientation of the drone to enable safe platooning. To enable the safe operation of the drone, Matthan~\cite{nguyen2017matthan} leverages the RF signals to detect if the drone is entering the private zone for safety. These industrial and complicated drones have been leveraged for different kinds of applications. For example, DroneScale~\cite{nguyen2020dronescale} can estimate the drone's loads using RF sensing. SkyCore~\cite{moradi2018skycore} and SkyRAN~\cite{chakraborty2018skyran} are designed to assist network access in disaster areas. RFly~\cite{ma2017drone} is designed to use a flying drone to extend the RFID's communication range.

In comparison to the large industrial drones, the small drones (e.g., nanodrone or quadrotor) are more flexible and agile, which have been used for human-drone interaction~\cite{monajjemi2016uav,abtahi2017drone,ng2011collocated, obaid2016would}. For example, \cite{yau2020subtle} exploits the force interaction and IMUs for drone control. BitDrones~\cite{gomes2016bitdrones} are designed as a form of programmable matter for human-drone interaction. HoverBall~\cite{nitta2014hoverball} is designed as a ball-shaped nanodrone, which can hover and change its location as required. Authors in~\cite{monajjemi2016uav} design a human-drone interaction system by using arm-waving gestures to interact with drones. Since these nanodrones are small, we cannot instrument different sensors on them for ubiquitous sensing. Instead, multiple nanodrones can collaborate with each other to formulate a swarm to enhance its sensing ability. TurboTrack~\cite{ma20163d} is designed to track RFID-tagged drones or small robots. However, TruboTrack uses the customized reader to read tags outside of the ISM band, which constrains it from being widely deployed for ubiquitous sensing. \sysname only uses the commodity passive RFID tags and reader without modifying RFID communication protocol, thereby \sysname could enable ubiquitous sensing, which is an extension of the work-in-progress paper~\cite{sun2023enabling}.


\section{Background}
\label{sec:background}


\subsection{Primer on Commodity Passive RFID}

\textbf{Working principle.} The commodity passive RFID system consists of the reader and tag, which complies with EPC Gen2 standard using the slotted ALOHA protocol~\cite{o2006gen}. The RFID tags are battery-free and need to be illuminated by the external reader through RF signals. So, the reader should initiate the communication by sending constant continuous waves (i.e., CW) to activate the tags within its communication range. The tag will reply to the reader's query with its own RN16. Physically, the tag's chip will control its impedance to match or mismatch its antenna's impedance (i.e., ON-OFF keying modulation), such that '0' and '1' bits are backscattered to the reader. After the reader receives the RN16s from the activated tags, the reader should choose one tag to communicate with by sending the acknowledgment. The chosen tag will transmit its own EPC (i.e., the tag's ID) back to the reader, and the other tags should keep silent and wait for communication in the next round.

\noindent\textbf{Form factor.} The commodity passive RFID system has an asymmetric structure, which pushes the complicated design at the reader and keeps the tag as simple as possible. The RFID tag consists of the chip and antennas, which are printed on the substrate. So, the commodity passive RFID tags are small form factor (e.g., ALN-9662 tag with a size of 7x2 cm and Monza 4 tag with a size of 4.5x4.5cm), which can be attached to the nanodrone. The lightweight RFID tag will not affect the nanodrone's functionalities.

\noindent \textbf{Cost and flexibility.} The RFID tag is low-cost (i.e., around 5 cents per tag). However, the reader is expensive (i.e., $\$1000+$ per reader) and bulky. To proliferate commodity passive RFID-based sensing, mobile, and low-cost readers are desired and have been designed recently. For example, TSL-1128~\cite{tsl}, ALR-S350~\cite{alr}, and Phychips~\cite{phychips} can extend the function of smartphones with additional accessories to identify commodity RFID tags. Some papers show that our handheld smartphone can be turned into a commodity RFID reader directly such as TiFi~\cite{an2018cross} and \cite{ensworth2015every}. We believe that these smartphone-enabled RFID readers can proliferate ubiquitous RFID sensing and decrease the system cost.

\subsection{Principle of Nanodrone Swarm's Platooning}

\textbf{Nanodrone.} There are two kinds of drone manufacturers: helicopter and multirotor. The multirotor-based drones are usually equipped with four rotors to propel the drone's body such as nanodrone (i.e., quadrotor). Since the multirotor-based drone is easy to maintain balance and maneuver, most commodity drones including nanodrones are multirotors~\cite{nguyen2017matthan}. There are four basic movements for nanodrones: roll rotation, pitch rotation, yaw rotation, and altitude rotation, which can adjust the thrust force to propel the drone's body. Since the nanodrone is a mechanical system, its body shift, rotation flip, or vibration will change the backscattered signals as the RFID tag is attached to the nanodrone's body. To see this clearly, we do an experiment to show the variation of phase readings over time from an RFID-tagged nanodrone. 
$\\$\textbf{Safe Platooning.}  We will see that multiple nanodrones are usually formulating a swarm for sensing applications due to the limited ability of the individual nanodrone. Therefore, it is important to maintain a desired 3D geometry for the nanodrone swarm during platooning, which requires the nanodrones to be instrumented with different sensors. However, it is impossible to instrument additional sensors on the nanodrone due to its small size, lightweight, and limited hardware enhancement. So, commodity passive RFID-tagged nanodrone is an appropriate way to recognize the 3D geometry of nanodrone swarm for safe platooning.

\begin{figure}
  \centering
  \includegraphics[width=\linewidth]{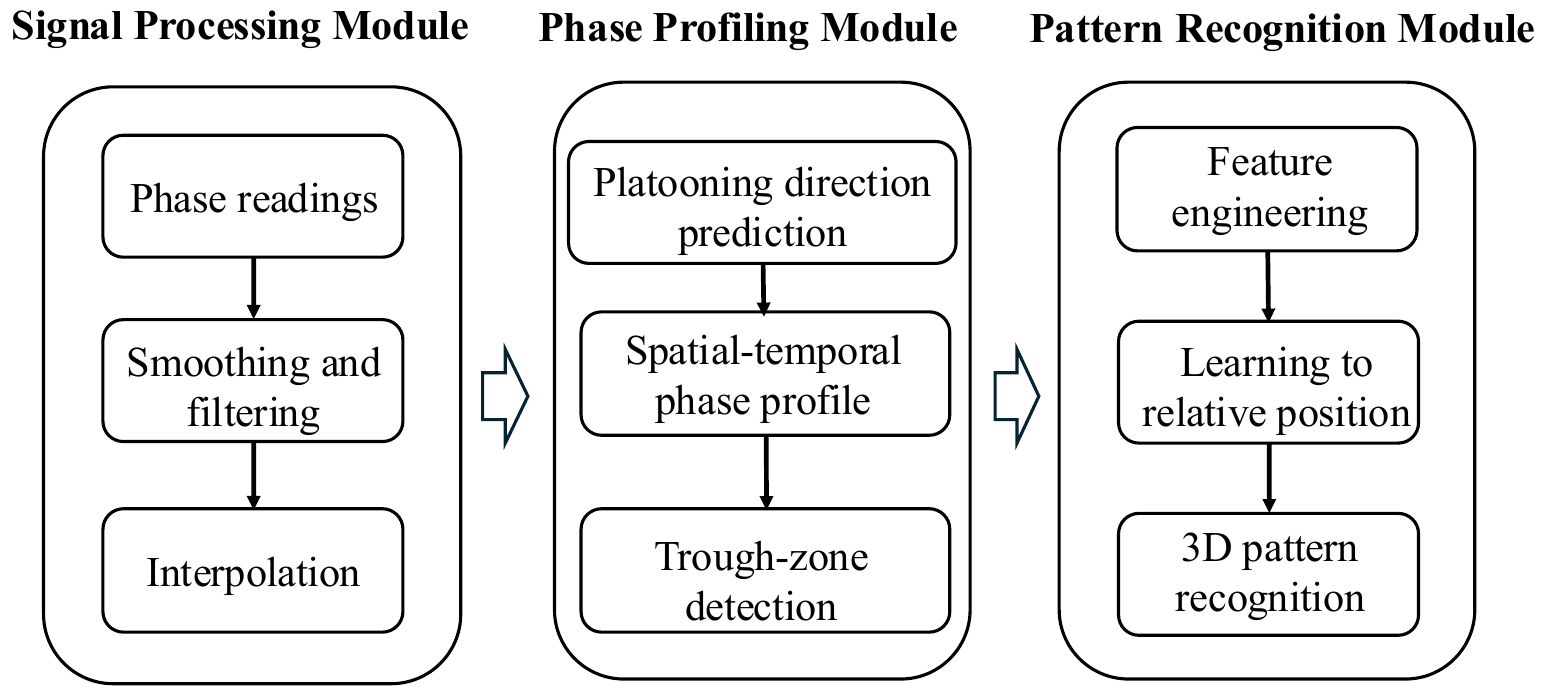}
    \caption{\sysname's workflow operation for relative localization of the nanodrones in the swarm, consisting of the signal processing module, phase profiling module, and 3D pattern recognition module.}
    \label{fig:overview:3d}
\end{figure}

\section{Overview}
\label{sec:overview}

\begin{figure*}
\centering
\captionsetup{width=0.23\textwidth}
\begin{minipage}{0.25\textwidth}
 \centering
  \includegraphics[width=\linewidth]{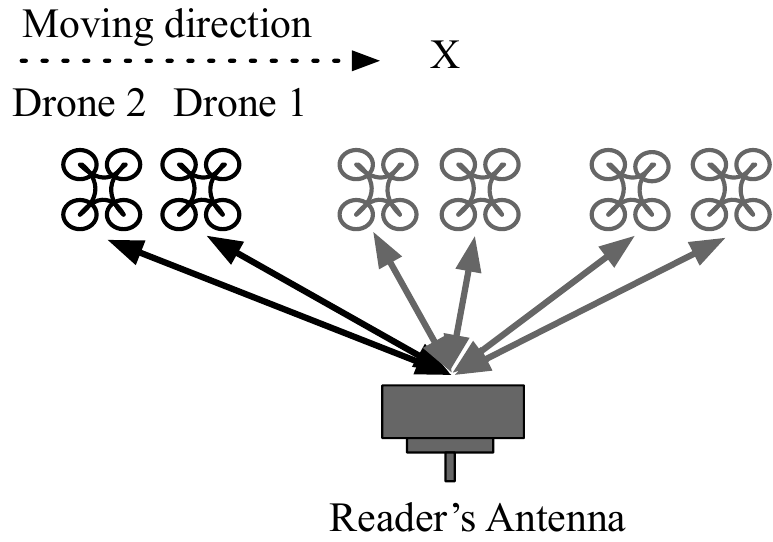} 
    \caption{Two nanodrones are moving from left to right in front of the reader's antenna along the x-axis.}
    \label{fig:two:drones:x}
\end{minipage}%
\begin{minipage}{0.25\textwidth}
  \centering
  \includegraphics[width=\linewidth]{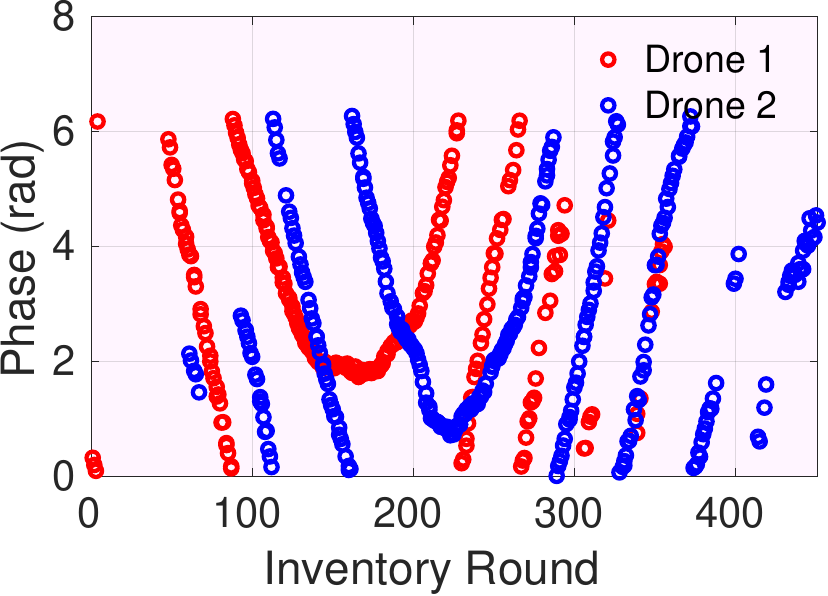} 
    \caption{Spatial and temporal phase profile from two nanodrones, indicating the relative position of them along the x-axis based on the time when the trough's lowest point has been achieved.}
    \label{fig:two:drones:x:phase}
\end{minipage}%
\begin{minipage}{0.25\textwidth}
 \centering
  \includegraphics[width=\linewidth]{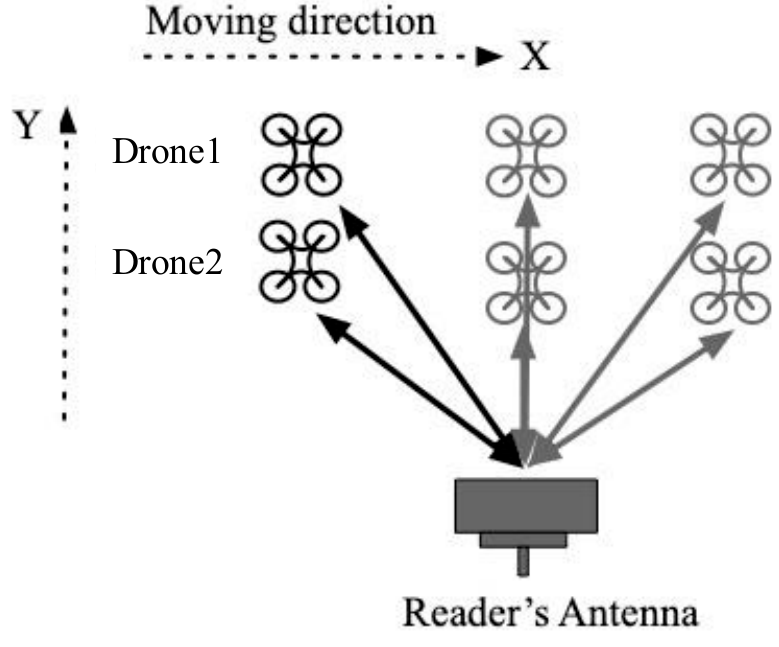}
    \caption{Two nanodrone are moving from left to right in front of the reader's antenna along the x-axis.}
    \label{fig:two:drones:y}
    \end{minipage}%
    \begin{minipage}{0.25\textwidth}
  \centering
  \includegraphics[width=\linewidth]{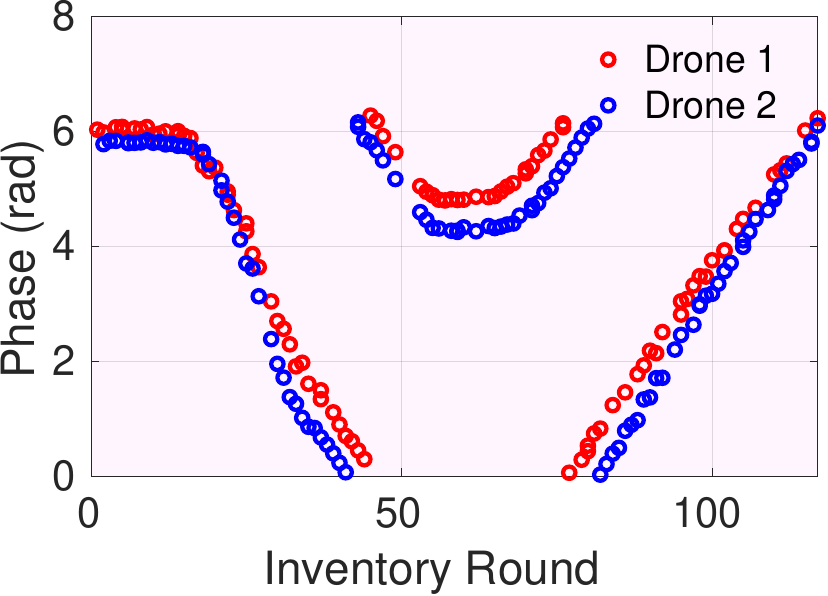} 
    \caption{Spatial and temporal phase profile from two nanodrones, indicating the relative position of them along the axis based on the value of trough's lowest point.}
    \label{fig:two:drones:y:phase}
\end{minipage} 
\end{figure*}

\textbf{Signal Processing Module.} The nanodrone swarm flies along the x, y, and z-axis during platooning. At the same time, the reader receives the backscattered signals from each RFID-tagged nanodrone. Before we forward the backscattered signals to the spatial-temporal phase profiling module, the extracted phase readings from the backscattered signals will be preprocessed with the Savitzky-Golay filter to eliminate the noisy points due to the nanodrone's body vibration. Moreover, the phase readings from each nanodrone in the swarm exhibit the same pattern, which can be harnessed to interpolate the phase readings. 

\noindent \textbf{Phase Profiling Module.} The preprocessed phase readings over time can be used to create the spatial-temporal phase profile. Specifically, we create three spatial-temporal phase profiles for each nanodrone in the swarm along the x, y, and z-axis during the platooning, using the colocated camera at the RFID reader. Then, we detect the trough zone on the spatial-temporal phase profile, which can characterize the relative position of the nanodrones in the swarm.


\noindent \textbf{Pattern Recognition Module.} After we obtain the spatial-temporal phase profiles along the x, y, and z-axis, we need to detect the trough zone in the profile for 3D nanodrone swarm pattern recognition. To do so, we design a bidirectional-LSTM model and use the spatial-temporal phase profile as the input to predict the relative positions of two nanodrones. After we obtain the relative positions of each pair of nanodrones across the x, y, and z axes, we can directly derive the 3D pattern of the nanodrone swarm.



\section{\sysname's Design}
\label{sec:design}

\subsection{Physical Principle of Relative Positioning}
\label{sub:sec:challenge}

To find the relative position of nanodrones in the swarm, we can use the linear relationship between the phase of backscattered signals and the reader-tag distance as follows:
\begin{equation}
\label{eq:phase}
    \theta = (\frac{2\pi * 2d}{\lambda} + \mu)\,\,mod\,\, 2\pi
\end{equation}
where $\theta$ is the phase of backscattered signals, $\lambda$ is the wavelength, $d$ is the distance between reader and tag  and $\mu$ indicates the phase shift due to the noise. As illustrated in the introduction section, the phase profile will exhibit a trough zone when the RFID-tagged nanodrone moves from left to right in front of the reader's antenna. Therefore, the relative position of two nanodrones along the x-axis can be estimated based on the time when the lowest point of the trough zone has been achieved. To see this clearly, we do experiments with a nanodrone swarm consisting of two nanodrones. The nanodrone 1 and 2 move from left to right in front of the reader's antenna along the x-axis as shown in Fig.~\ref{fig:two:drones:x}, where nanodrone 1 is at the right of nanodrone 2. So, nanodrone 1 is closer to the reader's antenna in comparison to nanodron 2. If we plot the phase profile of RFID-tagged nanodrone 1 and 2 as shown in Fig.~\ref{fig:two:drones:x:phase}, we can see two trough zones in the phase profiles of nanodrone 1 and 2. Moreover, the lowest point of the trough in the phase profile from nanodrone 1 will be at the left of the lowest point of the trough in the phase profile from nanodrone 2. This is because nanodrone 1 is closer to the reader's antenna in comparison to the nanodrone 2. Therefore, we can predict the relative position of nandorone 1 and 2 based on the time ordering of the lowest point of trough in their phase profiles.

After we figure out the relative position of two nanodrones along the x-axis, the problem becomes how we can predict the relative position of nanodrones in the swarm along the y-axis. To do so, we put two nanodrones in front of the reader's antenna with the same x coordinates and different y coordinates (e.g., the y coordinate of nanodrone 1 is smaller than nanodrone 2) as shown in Fig.~\ref{fig:two:drones:y}, where two nanodrones will move along the x-axis. Obviously, nanodrone 1 is closer to the reader's antenna. Then, we plot the phase readings of two nanodrones over time as shown in Fig.~\ref{fig:two:drones:y:phase}. Since two nanodrones have the same x coordinates, we cannot predict their relative positions based on the time ordering of the trough's lowest point in the phase profile. As we can see, the lowest points of trough from the phase profiles of nanodrone 1 and 2 will be achieved simultaneously. However, we can see that the lowest point of the trough in the phase profile from nandrone 1 is smaller than the lowest point of the trough in the phase profile from nanodrone 2. This is because nanodrone 1 is closer to the reader's antenna. Mathematically, in this scenario, the phase-changing rate over time can be expressed in the following equation as illustrated in Taggo~\cite{duan2021full}:
\begin{equation}
    R_p = \frac{d\theta}{dt} = \frac{4\pi}{\lambda}\frac{v^2t+vx_0}{\sqrt{(x_0+vt)^2+y_0^2}}
\end{equation}
where $v$ is the moving speed of the RFID-tagged object and $(x_0, y_0)$ indicates the coordinates of the object. As we can see, the smaller coordinate indicates the larger phase-changing rate, which is in accord with our above observation in the experiment. Therefore, we can predict the relative position of nanodrones along the y-axis based on the value of the trough's lowest point (or phase-changing rate) in the phase profile.

\begin{figure}
  \centering
  \includegraphics[width=0.5\linewidth]{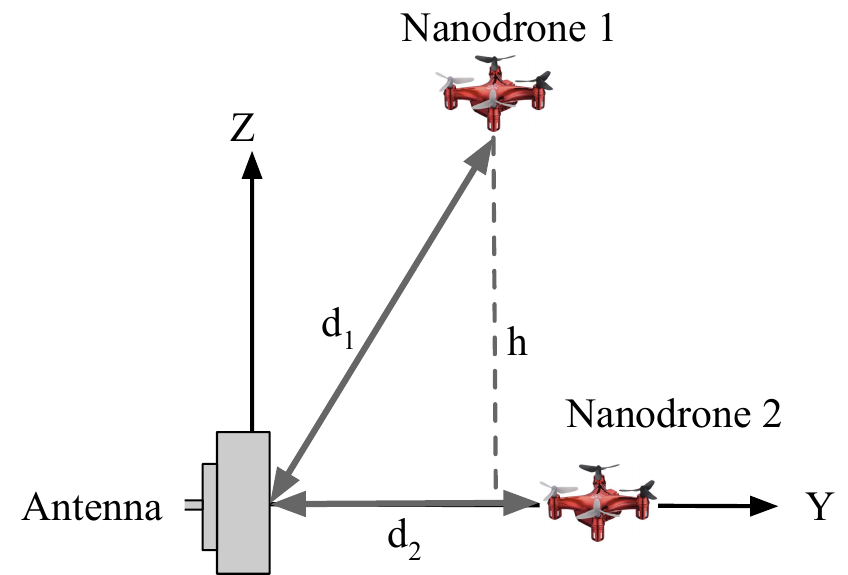}
    \caption{Nanodrone 1 is closer to the reader's antenna along the y-axis in comparison to naodrone 2. Both nanodrone 1 and 2 have the same x coordinate. However, the distance between nanodrone 1 and the reader's antenna (i.e., $d_1$) is larger than the distance between nanodrone 2 and the reader's antenna (i.e., $d_2$) due to the larger height (i.e., $h$) of nanodrone 1 along z axis.}
    \label{fig:height:example}
\end{figure}

Note that the above discussion assumes that the nandorones in the swarm have the same height along the z-axis. To predict the relative position of nanodrones in 3D space, we need to consider different heights of nanodrones along the z-axis. Fortunately, the time ordering for relative position estimation along the x-axis will not be affected by the different heights of the nanodrones along the z-axis, since the nanodrone closer to the reader's antenna along the x-axis will always achieve the lowest point of the trough in the phase profile earlier than the other nanodrones. However, the different heights of nanodrone along the z-axis will affect the estimation of nanodrone's relative position along the y-axis. As shown in Fig.~\ref{fig:height:example}, let's assume two RFID-tagged nanodrones have the same x coordinates. Nanodrone 1 has a smaller y coordinate and larger z coordinate in comparison to the nanodrone 2. In this case, if we plot the phase readings of two nanodrones and use above discussed approaches to estimate the nanodrone's relative positions along the y-axis, we will disorder their relative position along the y-axis based on the value of the trough's lowest point (or phase changing rate). This is because the distance (i.e., $d_1$) between nanodrone 1 and the reader's antenna is larger than the distance (i.e., $d_2$) between nanodrone 2 and the reader's antenna due to the larger height (i.e., $h$) of nanodrone 1. 

Since our goal is to estimate the relative position of nanodrones along the x, y, and z-axis in 3D space, the above discussion indicates that we cannot predict the relative position along the y-axis without considering the height of nanodrones along the z-axis. Moreover, we need to predict the relative position of the nanodrones along the z-axis. To this end, we profile the spatial-temporal phase along the x, y, and z-axis separately to achieve relative position estimation of the nanodrones in the swarm.

\subsection{Phase Profiling}
\label{sub:sec:phase}

When the nanodrone is platooning in the 3D space, all the nanodrones should fly across the x, y, or z axis simultaneously. As such, we should profile each nanodrone's phase readings across three axes. Therefore, it is important to predict the nanodrone swarm's platooning direction. 

\subsubsection{Platooning Direction Prediction}

Even though each nanodrone in the swarm is attached by an RFID tag, it is impossible to predict the nanodrone swarm's platooning direction across the x, y, and z axes. This is because the phase readings are not differentiable across the x, y, and z axes. To this end, we propose to deploy a camera colocated with the RFID reader for platooning direction prediction. Since the nanodrone swarm is platooning, each nanodrone in the swarm performs the same platooning operation. Even though we cannot predict each nanodrone's platooning direction, we can still identify the nanodrone swarm's platooning direction as long as there is one nanodrone in the camera's view. We regard the colocated camera and RFID reader to be the origin of 3D space. As such, the platooning direction can be obtained by estimating the nanodrone's flying direction in the camera's view. Then, we can easily match the time-stamped camera sensing data and the reader's readings. As such, based on the camera's sensing data, we are able to tell the platooning direction for the phase readings.

\begin{figure}
\centering
\captionsetup{width=0.23\textwidth}
\begin{minipage}{0.25\textwidth}
 \centering
  \includegraphics[width=\linewidth]{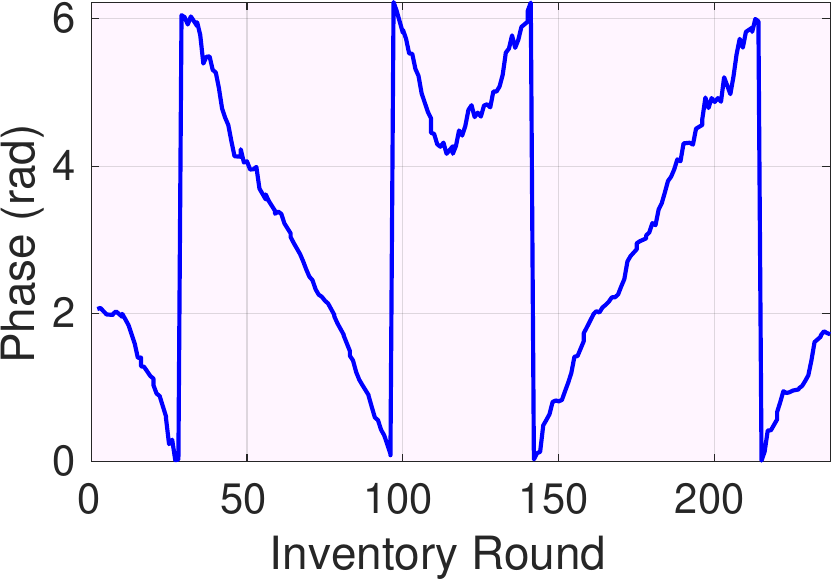}
    \caption{Raw phase readings over inventory rounds, which includes a trough zone from 100 to 150 inventory rounds.}
    \label{fig:raw:phase}
\end{minipage}%
\begin{minipage}{0.25\textwidth}
  \centering
  \includegraphics[width=\linewidth]{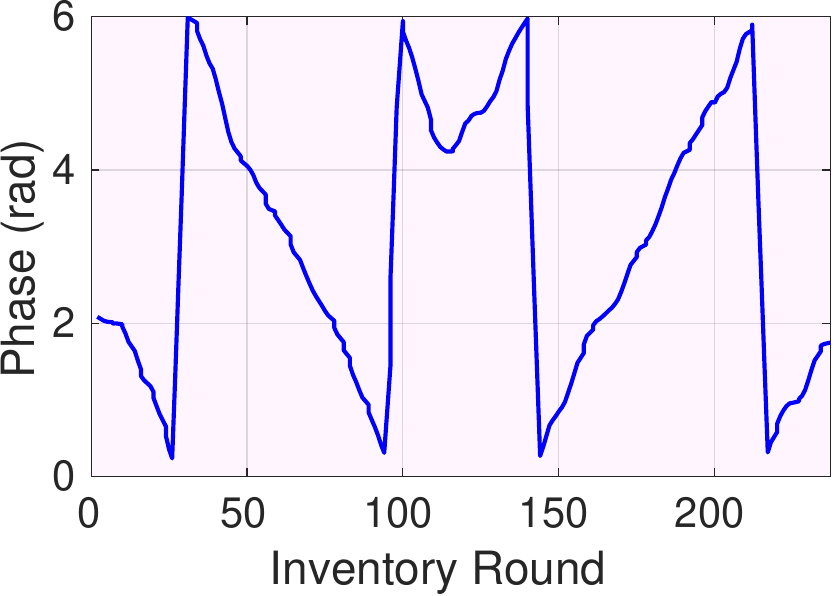}
    \caption{Phase readings over inventory rounds after we apply the Savitzky-Golay filter on raw phase readings to eliminate the noisy points.}
    \label{fig:phase:filtering}
\end{minipage} 
\end{figure}

\begin{figure*}
\centering
\captionsetup{width=0.31\textwidth}
\begin{minipage}{0.33\textwidth}
 \centering
  \includegraphics[width=\linewidth]{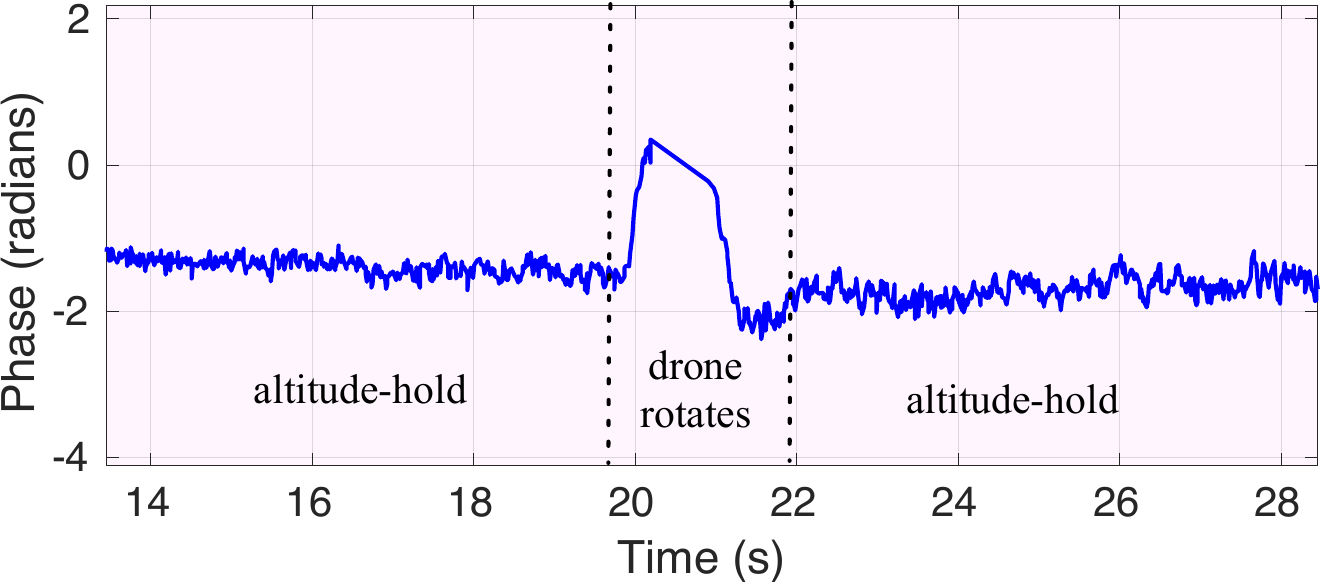}
    \caption{When the RFID-tagged nanodrone rotates/flips, the signal phase significantly changes. However, when the RFID-tagged nanodrone holds its altitude, the signal phase maintains constant.}
    \label{fig:flip:drone}
\end{minipage}%
\begin{minipage}{0.33\textwidth}
 \centering
  \includegraphics[width=0.8\linewidth]{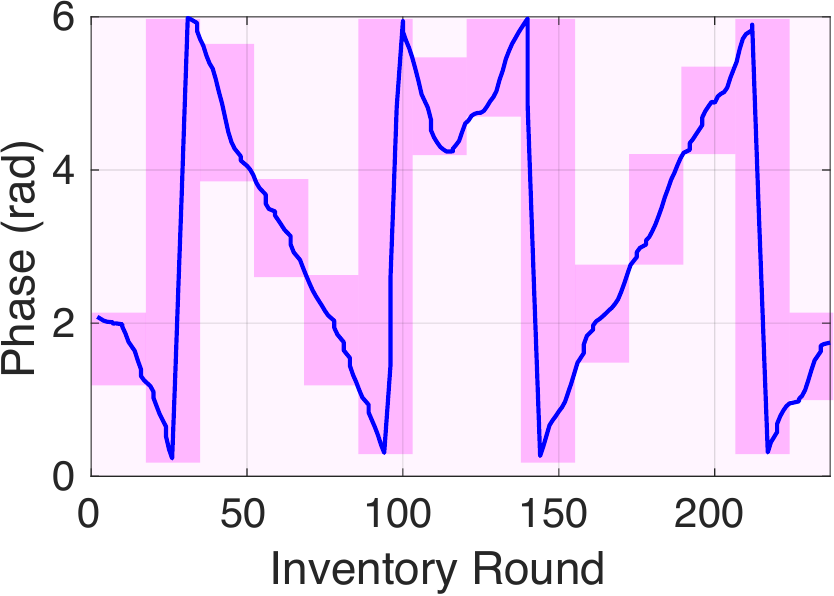}
    \caption{Trough zone detection with the sliding window on phase readings over inventory rounds.}
    \label{fig:phase:detection}
\end{minipage}%
\begin{minipage}{0.33\textwidth}
  \centering
  \includegraphics[width=0.8\linewidth]{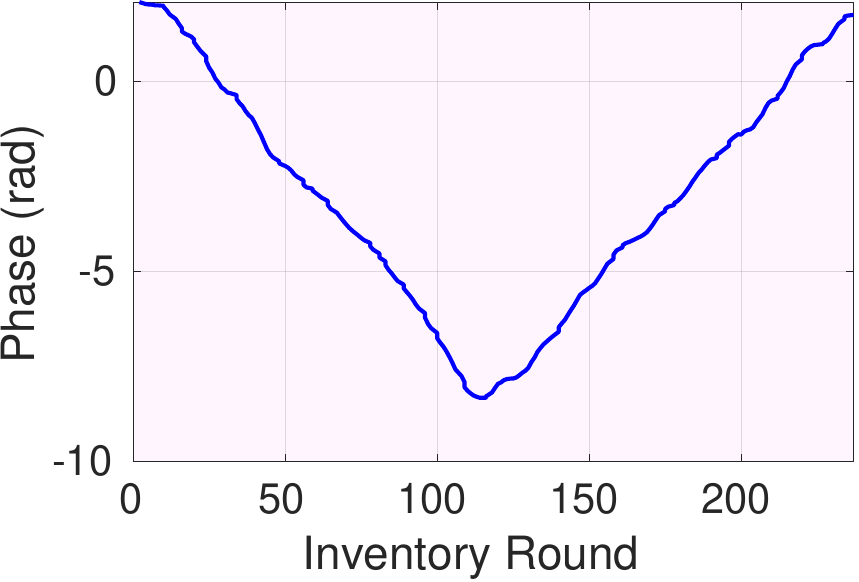}
    \caption{Phase readings over inventory rounds after we splice the adjacent parts together.}
    \label{fig:phase:splicing}
\end{minipage} 
\end{figure*}

\subsubsection{Spatial-Temporal Phase Profiling}

To predict the relative position of nanodrones in the swarm along the x, y, and z-axis, we will control the nanodrone swarm to fly along the x, y, and z-axis separately to obtain the time ordering of the trough's lowest point along the x, y, and z-axis in the spatial-temporal phase profile. When we move the nanodrone along a specific axis, the time ordering of trough's lowest point just depends on the nanodrone's coordinates along that axis.

However, the phase readings of backscattered signals are noisy due to the cluttered environment and nanodrone's body vibration. Thus, we need to preprocess the raw phase readings to obtain the clean spatial-temporal phase profile. To do so, we apply the Savitzky-Golay filter on the raw phase readings to mitigate the noisy points, which will help us detect the trough zone and recognize its lowest point for 3D nanodrone pattern recognition. As shown in Fig.~\ref{fig:raw:phase}, we show the raw phase readings over time when an RFID-tagged nanodrone is flying along the x-axis in front of the reader's antenna. After we apply the Savitzky-Golay filter to the phase readings, we can see the smooth phase readings over time as shown in Fig.~\ref{fig:phase:filtering}.

During the platooning, the nanodrone's body may not be stable and even flipped over  in the air, which introduce the artifacts to the phase readings and further affect the relative localization. As shown in Fig.~\ref{fig:flip:drone}, when the nanodrone holds its altitude (i.e., hovering in the sky), the signal phase is quite stable. When the nanodrone's body rotates, the signal phase changes significantly. This indicates that the nanodrone's body vibration does not affect the backscattered signals significantly. So, we can ignore the nanodrone's body vibration in our analysis. But, the nanodrone's body shift/rotation can significantly affect the backscattered signals, which need to be filtered out. To this end, we harness the platooning principle of the nanodrone swarm, where all the nanodrones perform the same platooning operation. As such, we can compare the phase reading pattern of all the nanodrones in the swarm to detect the abnormal vibration or rotation of some nanodrones and further filter out the phase readings that are affected. Specifically, we can use a sliding window and leverage Dynamic Time Warping (DTW) algorithm to compare the similarity of phase reading patterns between nanodrones in the swarm. When the similarity is larger than a threshold, both of two nanodrones get one vote. After we compare all the nanodrone pairs, we can filter out the phase readings from the nanodrones whose votes are larger than a threshold.

\subsubsection{Trough Zone Detection.} It is important to efficiently and accurately detect the trough zone in the spatial-temporal phase profile for relative localization, which will affect the performance of 3D nanodrone swarm pattern recognition. Our idea is to harness the phase reading pattern and further use a sliding window indicated by the shadowing area to go over the phase profile once as shown in Fig.~\ref{fig:phase:detection}. As we can see, the trough zone presents a characteristic pattern where the phase readings at the left and right of the trough zone are larger than the phase readings in the trough zone. Therefore, we can leverage the structure of the phase profile for trough zone detection. The trough zone can be found between two consecutive jumping points. Since we just go over the phase profile once, the overall computational complexity is $O(3(nN))$.


To detect the relative positions of the nanodrones in the swarm, we can detect the trough for relative localization. However, since the phase readings can be affected by the multipath and the noise, it is hard to detect the trough zone and even order troughs from two nanodrones' phase readings for relative localization. Therefore, we leverage the machine learning model to characterize the trough zone for relative localization of nanodrones in the swarm. 

\begin{figure*}
\centering
\captionsetup{width=0.31\textwidth}
\begin{minipage}{0.33\textwidth}
 \centering
 \includegraphics[width=0.8\linewidth]{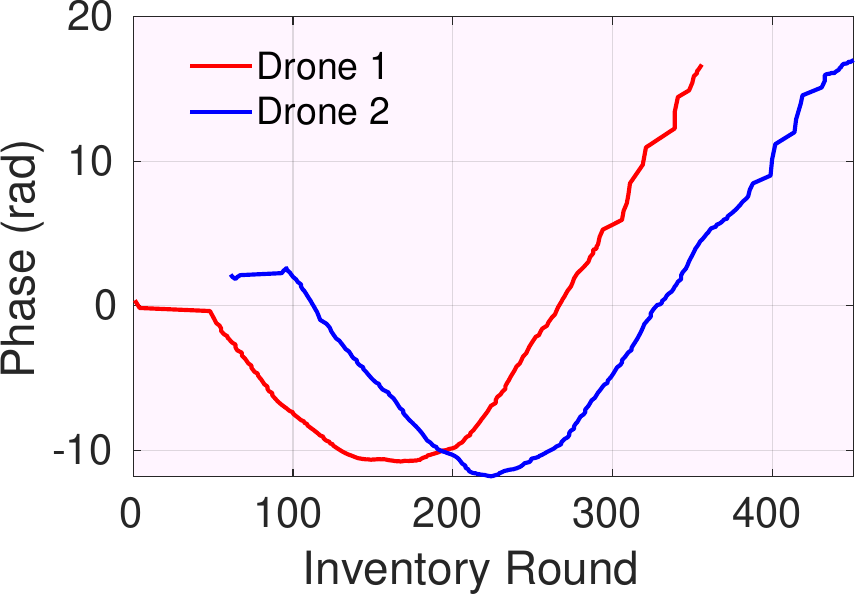}
    \caption{Spatial and temporal phase profile of two nanodrones flying along the x-axis.}
    \label{fig:fly:drone:x:phase}
\end{minipage}%
\begin{minipage}{0.33\textwidth}
  \centering
  \includegraphics[width=0.8\linewidth]{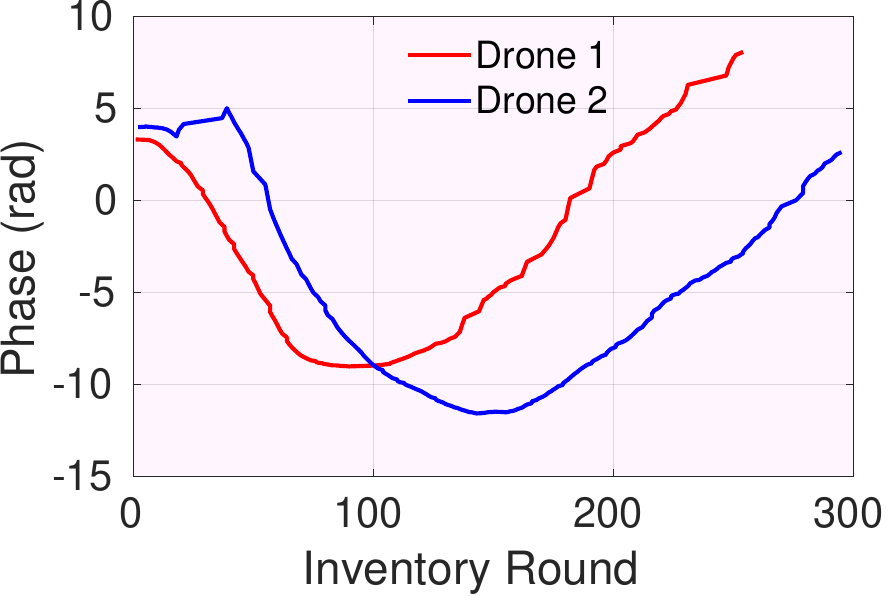}
    \caption{Spatial and temporal phase profile of two nanodrones flying along the y-axis.}
    \label{fig:fly:drone:y:phase}
\end{minipage} %
\begin{minipage}{0.33\textwidth}
  \centering
   \includegraphics[width=0.8\linewidth]{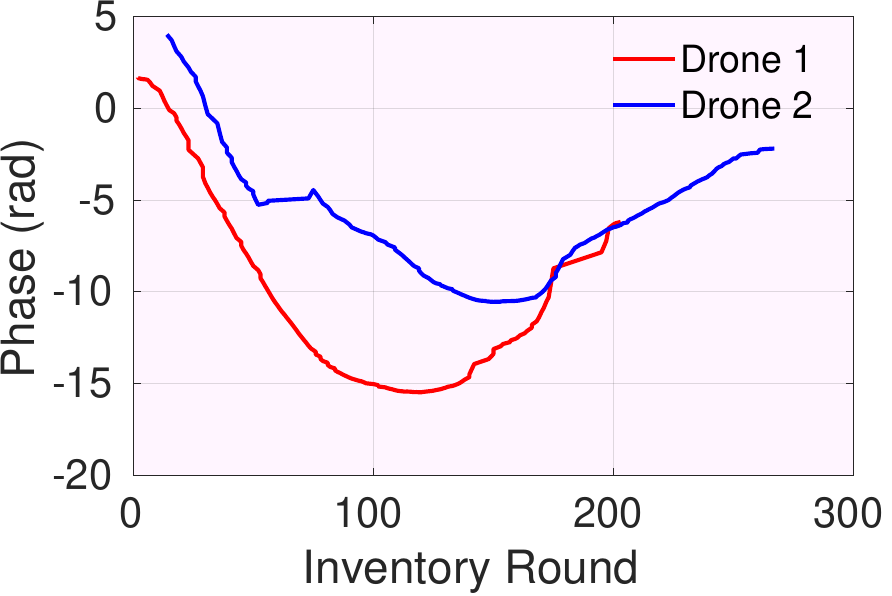}
    \caption{Spatial and temporal phase profile of two nanodrones flying along the z-axis.}
    \label{fig:fly:drone:z:phase}
\end{minipage} 
\end{figure*}

\noindent \textbf{Case study.} To demonstrate the feasibility of a spatial-temporal phase profile for relative localization, we platoon a nanodrone swarm with two drones (Drone 1 and 2) in an indoor environment along x, y, and z directions, where we assume $x_1>x_2$, $y_1>y_2$, and $z_1>z_2$. $(x_i,y_i,z_i),i=\{1,2\}$ are the coordinates of drone 1 and 2. We show the trough zone detected through our approach when the swarm platoons. Fig.~\ref{fig:fly:drone:x:phase}, Fig.~\ref{fig:fly:drone:y:phase}, and Fig.~\ref{fig:fly:drone:z:phase} show the spatial-temporal phase profile of two drones when they are flying along the x, y, and z axis respectively. As we can see, the lowest point in the trough zone of drone 2 exhibits a larger inventory round index in comparison to the trough zone of drone 1. This reveals that we can leverage the spatial-temporal phase profile to achieve relative localization. However, the phase readings are not resilient to the well-known environmental noise and multipath effect, which can potentially confuse the relative localization of the nanodrones in the swarm. Therefore, we plan to leverage the machine learning model trained on the spatial-temporal phase profile for the nanodrone relative localization. 

\subsection{3D Nanodrone Pattern Recognition with machine learning}
\label{sub:sec:pattern}

\subsubsection{Feature Engineering}

As we can see, the spatial-temporal phase profile consists of one trough zone and multiple discontinuous parts due to the modular operation. To achieve fast and accurate relative localization, we splice the discontinuous parts together with the trough zone as shown in Fig.~\ref{fig:phase:splicing} with the flowing equation~\cite{duan2021full}:
\begin{equation}
\label{eq:splicing}
\theta_i = \left\{\begin{matrix}
 \theta_i - \left \lceil \frac{\theta_i-\theta_{i-1}}{2\pi} \right \rceil 2\pi, &\theta_i-\theta_{i-1}>\pi \\ 
\theta_i + \left \lceil \frac{\theta_i-\theta_{i-1}}{2\pi} \right \rceil 2\pi, & \theta_i-\theta_{i-1}<-\pi\\ 
\theta_i,  & otherwise
\end{matrix}\right.
\end{equation}
where $\theta_i$ indicates the i-th phase value. To predict the relative position of one pair of nanodrones $a$ and $b$ in the nanodrones across the x-axis, when the nanodrone swarm platoons along the x-axis, we can profile their spatial-temporal profiles and further derive their trough zones indicated as $\boldsymbol{\theta}_a^x$ and $\boldsymbol{\theta}_b^x$, which can be used as the input of the machine learning model to predict the relative position of this pair of nanodrones. As a result, there are only two outputs for this machine-learning model. We indicate the first class as $c_1$ representing the relative position of nanodrones $a$ and $b$ along a specific axis. Similarly, for the relative position of two nanodrones across the y and z axes, we can obtain the trough zones of two nanodrones when they are platooning across the y and z axes. Then, we can predict the relative position of nanodrones in the swarm using our machine-learning model.

\begin{figure}
  \centering
  \includegraphics[width=0.5\linewidth]{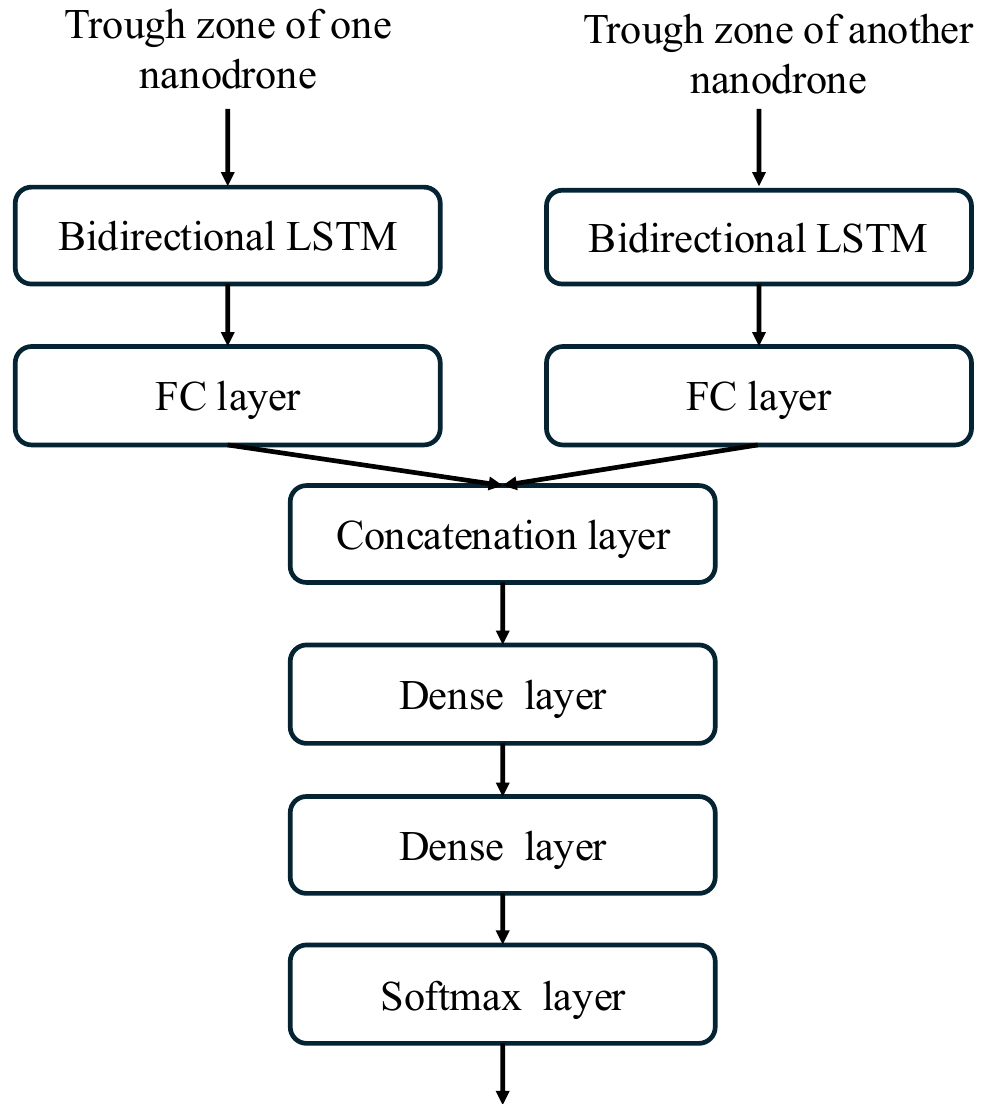}
    \caption{Learning to relative localization with bidirectional LSTM that can characterize the time-related trough zone over the inventory rounds.}
    \label{fig:lstm:net}
\end{figure}

\subsubsection{Learning to Relative Localization with LSTM} We propose a learning model that can identify the relative location of the nanodrones in the swarm. 

\noindent \textbf{Neural net architecture.} Our machine-learning model consists of two bidirectional LSTMs followed by a concatenation layer, one dese layer, one dropout layer, and one dense layer for relative position prediction of two nanodrones in the swarm. We use a bidirectional LSTM model that can characterize the time-related trough zone across the inventory rounds. As such, we can correctly predict the relative position of the nanodrones in the swarm based on their trough zones. Fig.~\ref{fig:lstm:net} shows our neural network architecture for relative position prediction. The output of this machine-machine learning model is the prediction of the relative position of two nanodrones in the swarm.

\noindent \textbf{Pattern recognition.} To recognize the relative position of the nanodrones in the swarm, we can use the well-trained neural network model to predict the relative position of each pair of nanodrones in the swarm across the x, y, and z axis. As such, we can obtain the nanodrone swarm's pattern in 3D space after all pairs of the nanodrones's relative positions are predicted. Note that our bidirectional LSTM can characterize the time-related trough zones across the inventory rounds to accurately predict the relative position of the nanodrones in the swarm.



\begin{figure}
\centering
\captionsetup{width=0.22\textwidth}
\begin{minipage}{0.25\textwidth}
 \centering
  \includegraphics[width=0.8\linewidth]{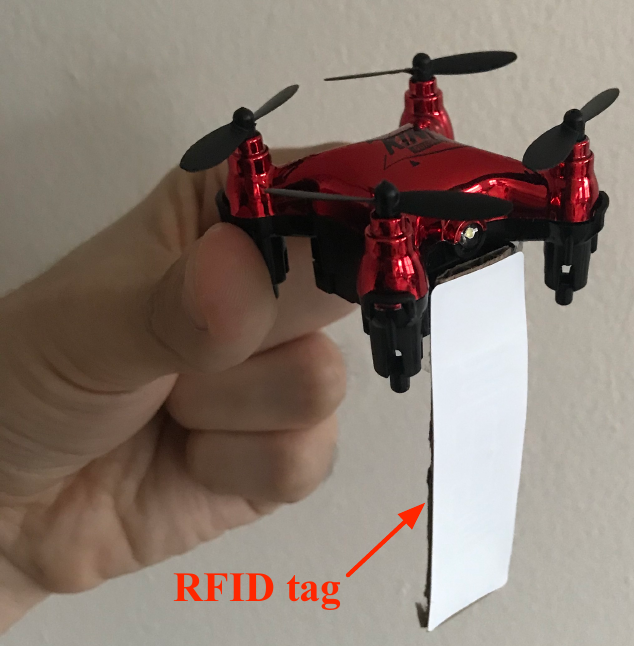}
    \caption{Commodity passive RFID tagged nanodrone.}
    \label{fig:tagged:drone}
\end{minipage}%
\begin{minipage}{0.25\textwidth}
  \centering
  \includegraphics[width=0.8\linewidth]{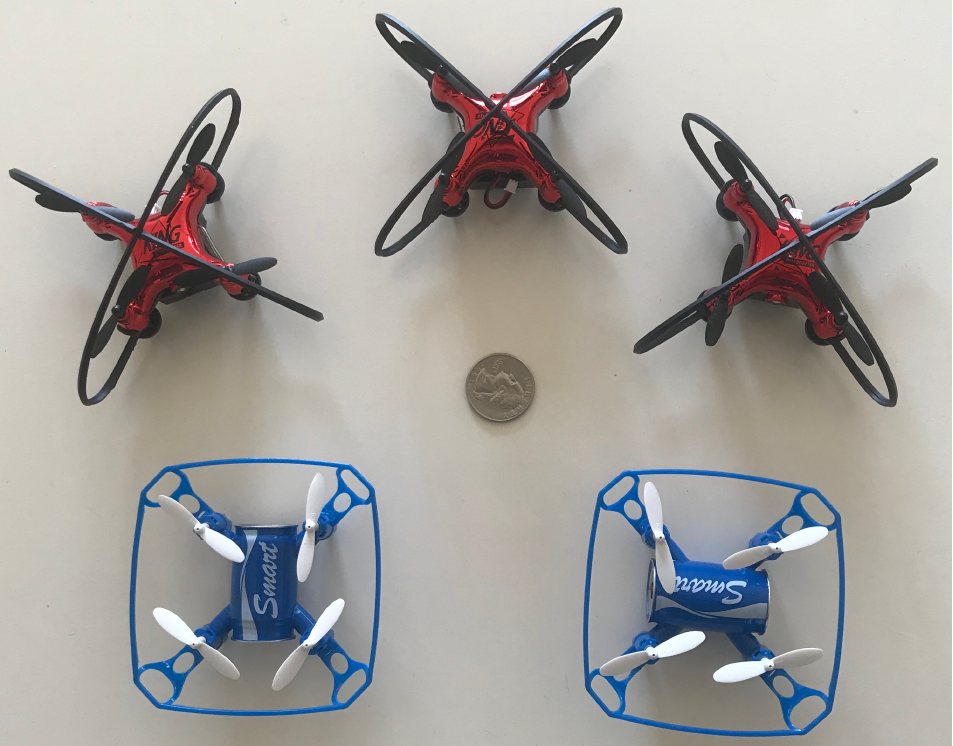}
    \caption{\textbf{Nanodrone swarm.} There are five nanodrones in the swarm. Three Holyton HT02 Mini Drones~\cite{holy} and two Masefu Mini Drones~\cite{masefu}}
    \label{fig:swarm}
\end{minipage} 
\end{figure}

\section{Implementation and evaluation}
\label{sec:imp:eva}

\noindent \textbf{RFID Reader and camera.} We adopt the USRP N210 radio as the RFID reader developed in the prior work~\cite{kargas2015fully}, which is instrumented with an SBX daughter board to interrogate the tags. It is compatible with FCC regulation using the EPC Gen2 standard for UHF RFID communication at a frequency band between 902-928MHz. The reader instrumented with the directional antennas will activate and interrogate the tags using the slotted ALOHA protocol. During experiments, we will just extract the channel state information (specifically, the phase readings over time) for nanodrone relative localization, which will be used for nanodrone swarm pattern recognition. The RFID reader is colocated with a camera (i.e., Intel RealSense Depth Camera D435
) for nanodrone swarm's platooning direction estimation.

\noindent \textbf{RFID Tag.} We use the general-purpose commodity passive RFID tags (e.g., Alien Squiggle RFID tag ALN-9640~\cite{aln_9640}, ALN-9762~\cite{aln_9762} and ALN-9662~\cite{aln_9662}) to evaluate \sysname's performance. Each tag is low-cost with the price of around 5 cents, which will enable ubiquitous sensing. To minimize the effect of the tag's orientation on the performance of 3D nanodrone swarm pattern recognition, each tag is vertically polarized on the nanodrone~\cite{luo20193d} as shown in Fig.~\ref{fig:tagged:drone}. We can also use circularly polarized RFID tags to eliminate the effect of the tag's orientation.

\noindent \textbf{Nanodrone Swarm.} The nanodrone swarm consists of multiple nanodrones. In our prototype, we use up to five nanodrones (e.g., three Holyton HT02 Mini Drones~\cite{holy} and two Masefu Mini Drones~\cite{masefu}) to formulate a nanodrone swarm as shown in Fig.~\ref{fig:swarm}. Each individual nanodrone has three-speed modes: low-speed mode of around $0.15m/s$, medium-speed mode of around $1m/s$, and high-speed mode of around $2m/s$.  During the experiments, each individual nanodrone in the swarm is controlled by the remote controller. The adjacent nanodrones will be separated by half a wavelength away to avoid the tag-tag coupling effect for safe platooning. 

 \begin{figure}
\centering
\captionsetup{width=0.22\textwidth}
\begin{minipage}{0.25\textwidth}
  \centering
\includegraphics[width=0.8\linewidth]{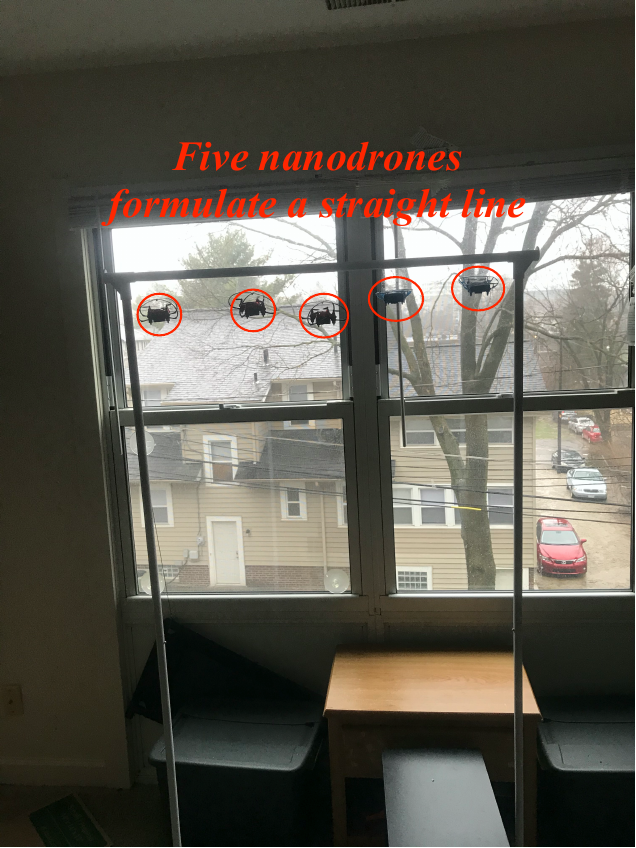}
     \caption{An example of five nanodrones formulating a straight-line pattern, where the garment rack is used for reference.}
    \label{fig:ex:line}
\end{minipage}%
\begin{minipage}{0.25\textwidth}
  \centering
 \includegraphics[width=0.8\linewidth]{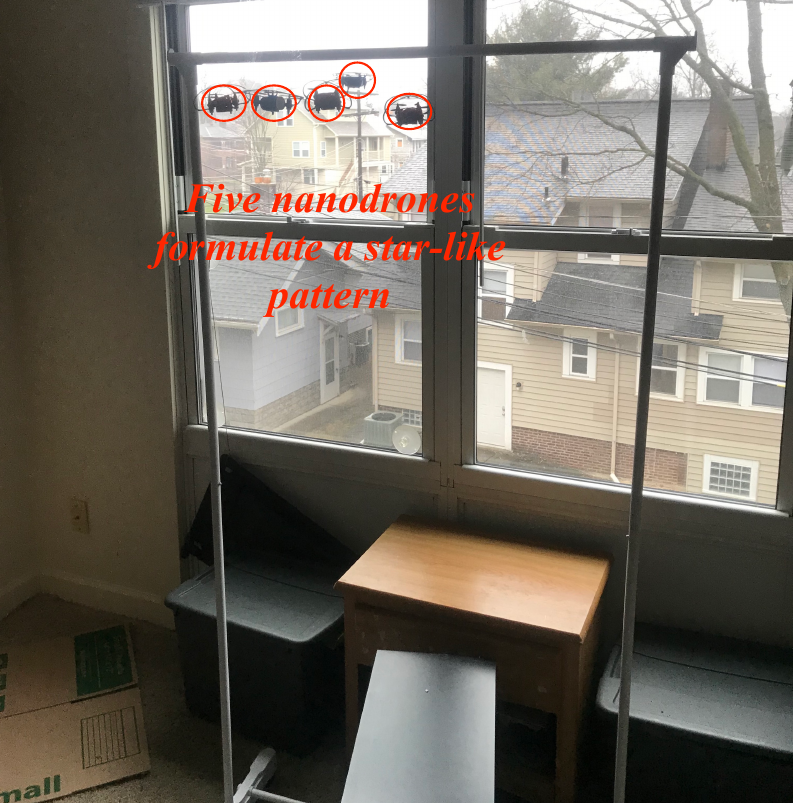}
    \caption{An example of five nanodrones formulating a star-like pattern, where the garment rack is used for reference.}
    \label{fig:ex:star}
\end{minipage} 
\end{figure}

\begin{figure*}
\centering
\captionsetup{width=0.31\textwidth}
\begin{minipage}{0.33\textwidth}
 \centering
  \includegraphics[width=\linewidth]{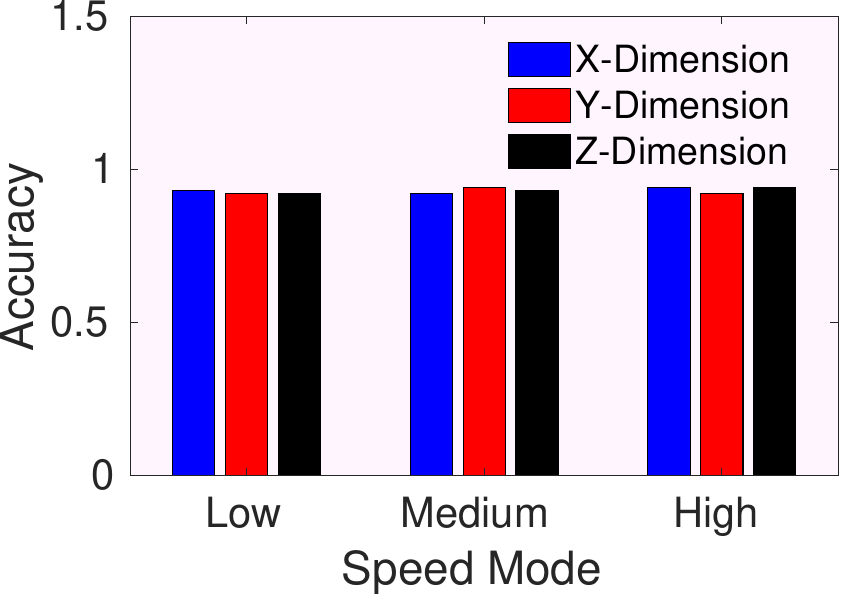}
    \caption{Relative localization accuracy along the x, y, and z-axis for nanodrone swarm with different speed.}
    \label{fig:relative:speed}
\end{minipage}%
\begin{minipage}{0.33\textwidth}
  \centering
  \includegraphics[width=\linewidth]{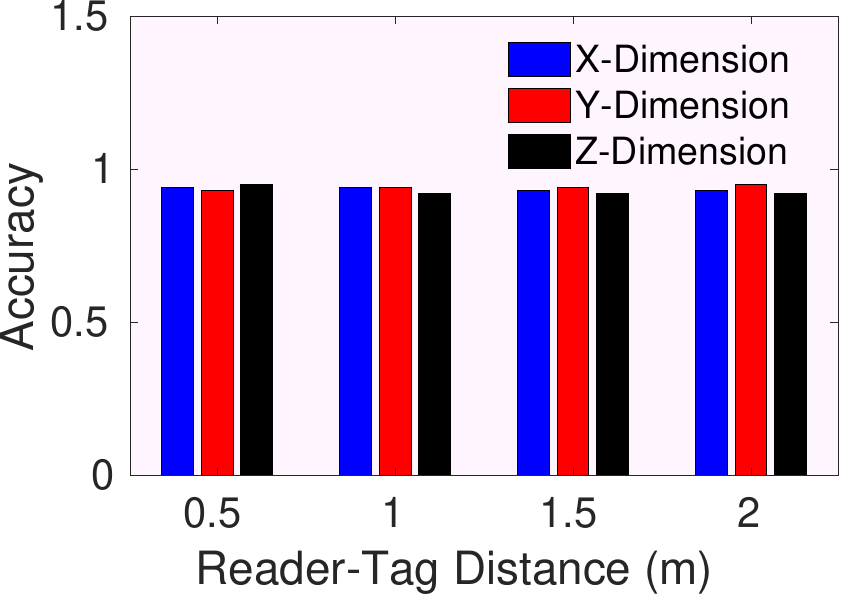} 
    \caption{Relative localization accuracy along the x, y, and z-axis across different reader-tag distance.}
    \label{fig:relative:distance}
\end{minipage} %
\begin{minipage}{0.33\textwidth}
  \centering
  \includegraphics[width=\linewidth]{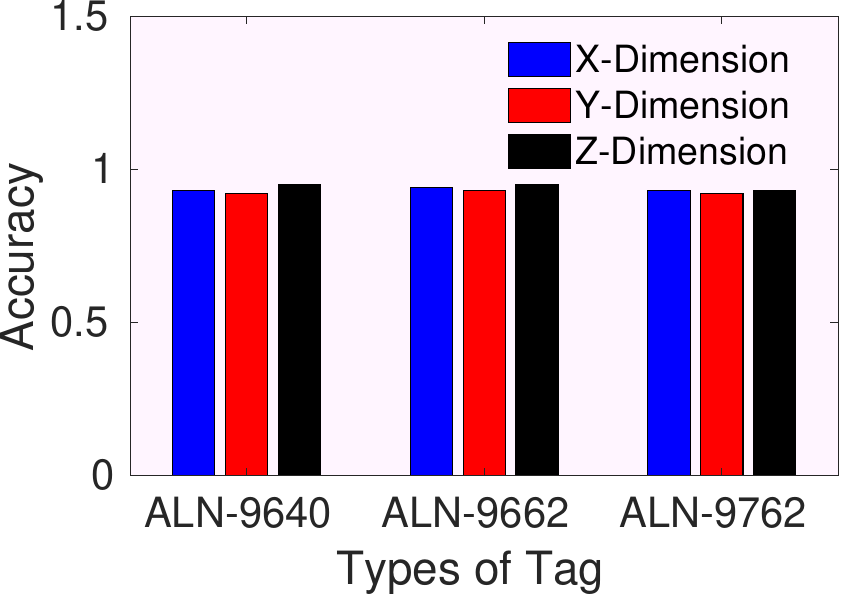} 
    \caption{Relative localization accuracy along the x, y, and z-axis for different tag types.}
    \label{fig:relative:types}
\end{minipage} 
\end{figure*}
\noindent \textbf{Experimental Settings.} The reader will connect with the PC host (i.e., an HP laptop running Ubuntu 16.04 operating system) through Ethernet cables. We use USRP N210 as the RFID reader to interrogate the tags and extract the backscattered channel between each RFID-tagged nanodrone and the reader's antenna. The reader is responsible for collecting the backscattered signals, which will be further forwarded to the PC host and processed with MATLAB in the PC host. After we obtain the backscattered channel, we leverage the well-trained machine-learning model to predict the relative position of the nanodrones in the swarm. To train the model, we fly two nanodrones in the indoor environment and record the phase readings from two nanodrones as the training dataset.

We evaluate the performance of \sysname in indoor and outdoor environments (e.g., apartment rooms and outdoor open spaces) with up to five nanodrones platooning in different patterns as shown in Fig.~\ref{fig:ex:line} and Fig.~\ref{fig:ex:star}. The apartment room is a rich scattering indoor environment with different furniture around. The outdoor environment is an open space. The distance between the nanodrone swarm and the reader's antenna is around 1.5 meters within the reader's communication range by default. We control them through the remote controller. In default, we control the nanodrone swarm to fly in the low-speed mode. We collect the phase profile measurements from the platooned nanodrone swarm in different indoor environments, where the 80 percentile is used to train the LSTM-based neural net, and the residuals are used to test its relative localization accuracy. In the results section, we measure the impact of different system settings on \sysname's performance such as tag type, reader-tag distance, flying speed, and different swarm patterns. The relative localization accuracy for a pair of nanodrones in the swarm is defined as the ratio of the correct predictions and total platooning trials. Then, we define the relative localization accuracy for the nanodrone swarm as the average relative localization accuracy for all pairs of nanodrones in the swarm.

\section{Experimental Results}
\label{sec:results}

\subsection{Accuracy of Platooning Direction Estimation}
The platooning direction estimation accuracy can significantly affect the relative localization across three axes. 

\noindent \textbf{Method.} So, we fly nanodrone swarm in the indoor environment by regarding the camera as the origin. The platooning direction is estimated through the nanodrone's flying direction in the camera's view. We report the accuracy of platooning direction estimation using the ratio of the correct estimation trials and the total flying trials. The ground-truth platooning direction can be controlled by the controller who interacts with the nanodrone swarm.

\begin{table}
  \centering
  \includegraphics[width=0.5\linewidth]{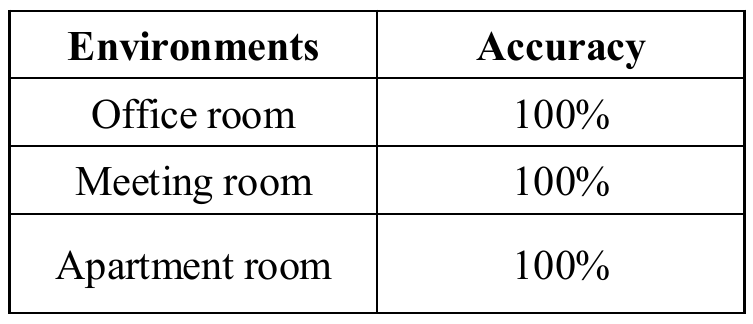}
    \caption{Accuracy of nanodrone swarm platooning direction estimation across different indoor environments.}
    \label{table:platooning:direction}
\end{table}

\noindent \textbf{Result.} Table~\ref{table:platooning:direction} shows the accuracy of the platooning direction estimation in the different indoor environments. As we can see, we can correctly estimate the nanodrone swarm's platooning direction across different indoor environments. This is because the camera can correctly predict at least one nanodrone's flying direction in the swarm.

\subsection{Relative Localization Accuracy}

\subsubsection{Impact of Nanodrone Swarm's Speed} The flying speed of the nanodrone swarm is an important factor for 3D nanodrone swarm pattern recognition since the number of phase readings depends on the tag reading rate.

\noindent \textbf{Method.} To measure the impact of the swarm's speed, we fly a nanodrone swarm with different speed modes (i.e., low, medium, and high) along the x, y, and z-axis. We report the relative localization accuracy along the x, y, and z-axis. The nanodrone swarm will fly within the reader's communication range.

\noindent\textbf{Result.} Fig.~\ref{fig:relative:speed} shows the relative localization accuracy along the x, y, and z-axis, when we fly the nanodrone swarm with low, medium, and high-speed mode. As we can see, the median accuracy of relative localization is around 0.95, 0.93, and 0.93 for low, medium, and high-speed modes respectively. Therefore, the nanodrone swarm's speed will not degrade the performance of relative localization due to the high tag reading rate (i.e., 1000+ tags per second~\cite{impinj}) of the commodity RFID reader. We also see the high relative localization accuracy for nanodrone swarm flying along the x, y, and z-axis. This is because we use the same relative localization approach to decide the relative position of nanodrones along the x, y, and z-axis, thereby the relative localization across the x, y, and z-axis is independent.

\subsubsection{Impact of Reader-Tag Distance} Since the backscattered signals are suffering from propagation loss, the distance between the reader's antenna and RFID-tagged nanodrone will affect the performance of nanodrone swarm pattern recognition. 
$\\$ \textbf{Method.} To measure the impact of reader-tag distance on the relative localization, we fly a nanodrone swarm at different distances from the reader's antenna. The distance of nanodrone swarm to the reader's antenna is defined as the average distance of each individual nanodrone in the swarm to the reader's antenna. More importantly, the nanodrone swarm will fly within the reader's communication range and there are other reflectors around (e.g., furniture and walls in the apartment room). $\\$ \textbf{Result.} Fig.~\ref{fig:relative:distance} shows the relative localization accuracy of nanodrone swarm along the x, y, and z-axis across different reader-tag distances. As we can see, the relative localization accuracy over the x, y, and z-axis for the specific reader-tag distance is similar, since we use the same relative localization approach. Moreover, the median relative localization accuracy for different reader-tag distances is similar (i.e., around 0.93), whereby the reader-tag distance will not affect the relative localization as long as the nanodrone swarm flies within the reader's communication range. This is because we leverage the shape of the trough zone for relative localization, which is independent of the signal propagation loss.

\subsubsection{Impact of Different Tag Types} There are different general-purpose commodity passive RFID tags on the market. In our experiment, we use three commodity passive RFID tags to measure the tag diversity on relative localization accuracy.

\noindent\textbf{Method.} We attach different RFID tags to the nanodrone in each experiment. The RFID-tagged nanodrone swarm will fly along the x, y, and z-axis for relative localization within the reader's communication range. Note, that the lightweight and battery-free commodity RFID tag will not affect the nanodrone swarm's fly.

\noindent\textbf{Result.} Fig.~\ref{fig:relative:types} shows the relative localization accuracy along x, y and z axis across different tag types. The median relative localization accuracy for ALN-9640, ALN-9662, and ALN-9672 RFID tags is around 0.92, 0.93, and 0.91, which indicates the tag types will not affect the relative localization accuracy as long as the reader can read RFID tags. For the specific RFID tag, the relative localization accuracy is similar across the x, y, and z-axis, since the relative localization algorithm is independent on different axes. Therefore, the tag diversity will not affect the performance of relative localization, which will potentially advance \sysname's ubiquitous sensing for 3D nanodrone swarm pattern recognition.

\begin{figure}
  \centering
  \includegraphics[width=\linewidth]{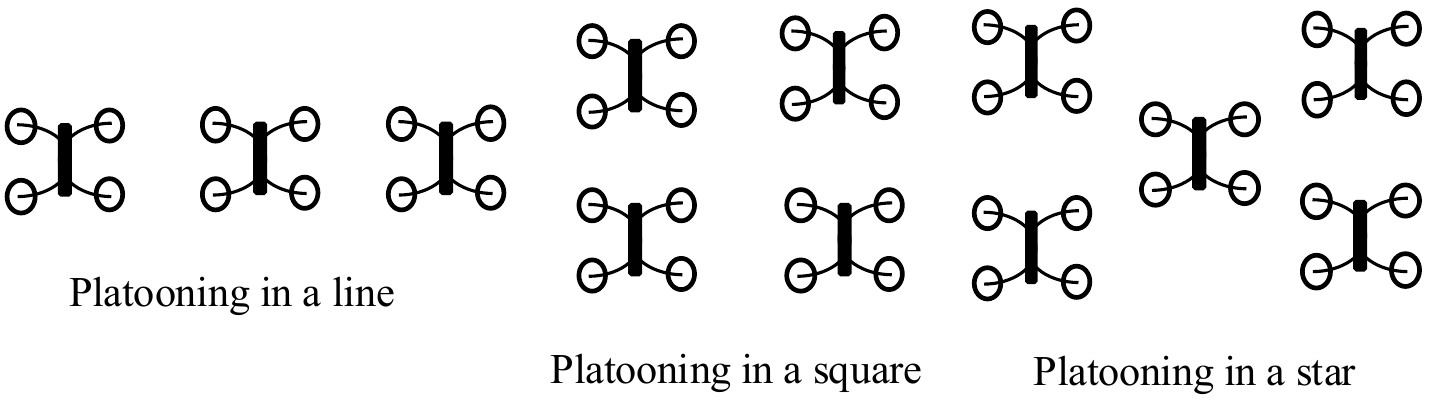}
    \caption{Platooning pattern examples of the nanodrone swarm.}
    \label{fig:pattern}
\end{figure}

\subsection{Accuracy of 3D Geometry Estimation for Nanodrone Swarm}


\subsubsection{Impact of 3D Geometry of Nanodrone Swarm Patterns} We will see the performance of \sysname on different 3D geometry of nanodrone swarm patterns since the nanodrone swarm can formulate different 3D geometry during flying.

\noindent \textbf{Method.}  To measure the effect of 3D geometry of nanodrone swarm patterns, we fly the nanodrone swarm consisting of three or five nanodrones with different patterns such as star pattern, line pattern, rectangular pattern, and triangular pattern. Fig.~\ref{fig:pattern} shows an example of the nanodrone swarms with different platooning patterns. We estimate the 3D geometry of nanodrone swarm pattern based on the predicted relative positions of nanodrones in the swarm. 

\begin{figure}
\centering
\captionsetup{width=0.23\textwidth}
\begin{minipage}{0.25\textwidth}
 \centering
  \includegraphics[width=\linewidth]{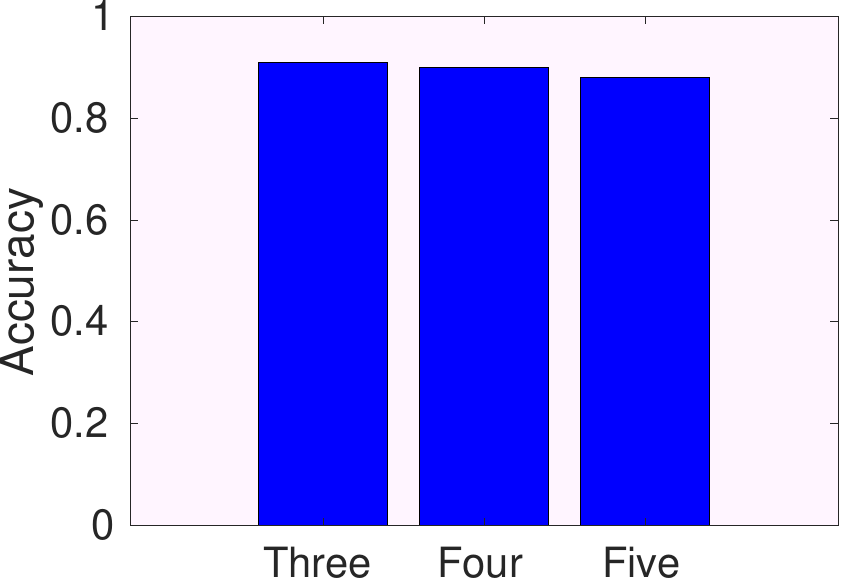}
    \caption{Accuracy of 3D geometry estimation for nanodrone swarm consisting of three, four, or five nanodrones.}
    \label{fig:geometry:swarm:size}
\end{minipage}%
\begin{minipage}{0.25\textwidth}
  \centering
  \includegraphics[width=\linewidth]{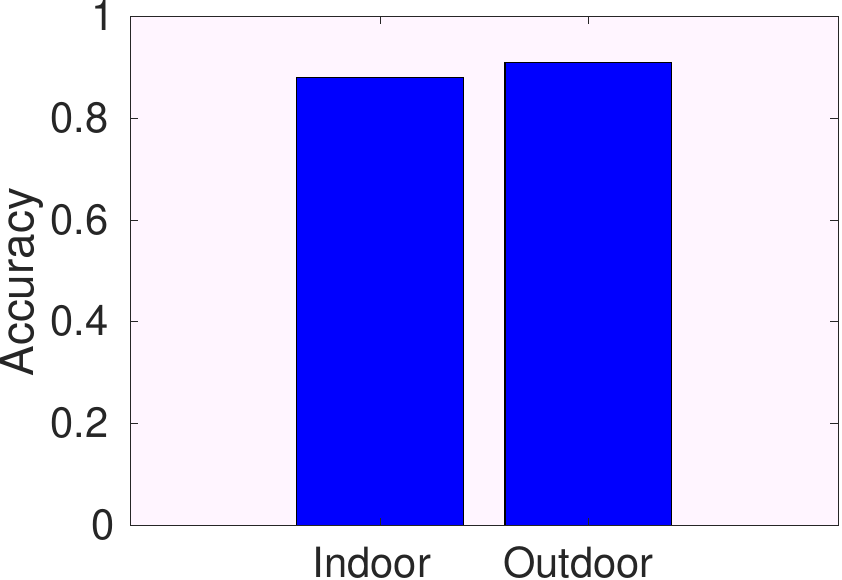}
    \caption{Accuracy of 3D geometry estimation for nanodrone swarm flying in the indoor or outdoor environment.}
    \label{fig:geometry:inout:door}
\end{minipage} 
\end{figure}
\noindent\textbf{Result.} Fig.~\ref{fig:geometry:swarm:size} shows the accuracy of 3D geometry estimation for nanodrone swarm consisting of three, four, and five nanodrones respectively. As we can see, the accuracy for three, four, and five nanodrones in the swarm is 0.92, 0.91, and 0.88 respectively, which is lower than the relative localization accuracy across different axes. This is because 3D geometry estimation for nanodrone swarm patterns depends on the relative localization. Moreover, the 3D geometry estimation for nanodrone swarm is not affected by the number of nanodrones in the swarm, as long as the reader can read all the RFID-tagged nanodrones. We also discuss the nanodrone swarm consisting of a large number of nanodrones in the discussion section ($\S \ref{sec:dis}$).

\subsubsection{Impact of Environments} The indoor environment (e.g., apartment room) is known as the rich scattering environment with many reflectors (e.g., furniture and walls), while the outdoor environment mainly exhibits line-of-sight signal propagation.

\noindent \textbf{Method.} To measure the effect of indoor and outdoor environments on the performance of nanodrone swarm pattern recognition, we fly nanodrone swarm in the indoor environment (e.g., an apartment room) and outdoor environment (e.g., an open space).

\noindent \textbf{Result.} Fig.~\ref{fig:geometry:inout:door} shows the accuracy of 3D geometry estimation for nanodrone swarm in indoor and outdoor environments. As we can see, the accuracy of 3D geometry recognition for nanodrone swarm in indoor and outdoor environments is 0.90 and 0.92 respectively, which is close to each other. Even though the indoor environment shows a rich scattering characteristic compared to the outdoor environment, the environmental diversity will not affect nanodrone swarm pattern recognition. This is because \sysname is relying on the relative localization based on the shape of the spatial-temporal phase profile, which does not require the accurate position of the nanodrones.

\section{Discussion, Limitations, and Future Work}
\label{sec:dis}


\noindent \textbf{Fine-Grained 3D Localization and Tracking.} \sysname can predict the relative position of the nanodrones to recognize the 3D geometry of the nandorone swarm pattern. Since we attach a commodity passive RFID tag on each nanodrone in the swarm, we can also accurately localize and track each nanodrone in the swarm. However, this is a challenging problem due to the cluttered indoor environment, which cannot be simply solved with a commodity passive RFID system. So, the state-of-the-art RFID-based localization and tracking techniques either leverage space diversity using a reader instrumented with multiple antennas or multiple readers (e.g., PinIt~\cite{wang2013dude} and ) or use a tag that can work with wideband (e.g., TurboTrack~\cite{luo20193d} and RFind~\cite{ma2017minding}). More importantly, localizing or tracking multiple RFID-tagged nanodrones in the swarm will cause a large processing delay due to the time-division channel access protocol employed by the RFID system. Recently, we see some parallel decoding techniques~\cite{ou2015come, jin2020fireworks, xu2020embracing, hu2015laissez} have been developed to improve the performance of RFID system, which can be applied for simultaneous sensing in nanodrone swarm pattern recognition. However, all these parallel decoding systems can only technically support tens of tags, which cannot deal with a large number of nanodrones in the swarm.

\noindent \textbf{Flying Nanodrone Swarm in High Speed.} The nanodrone can fly at a very high speed. For example, Black Hornet Nano can fly at the speed of up to 13mph~\cite{black_drone}. The higher the speed the nanodrone flies, the higher the reading rate we require for the commodity passive RFID system. Note that the current commodity RFID reader can read $1000+$ tags per second~\cite{impinj}, which is sufficient to support nanodrone flying at high speed. Note that we can also leverage the parallel decoding techniques mentioned in the above subsection to improve the sensing performance of the RFID system, which can support the nanodrone swarm flying at very high speed. Fortunately, the relative localization of nanodrone in the swarm is estimated based on the time ordering of the trough's lowest point in the spatial and temporal phase profile. So, we can control the nanodrone swarm to fly along the x, y, and z-axis at a low speed for relative position estimation.

\noindent \textbf{Flying a Large Number of Nanodrones.} Our current prototype shows the good performance of \sysname with up to five nanodrones in the swarm. In real-world applications, we may fly a nanodrone swarm with more than five nanodrones, which will potentially increase the processing delay for nanodrone swarm pattern recognition due to the time-division channel access protocol used by the commodity passive RFID system. However, the commercial RFID reader (e.g., Impinj R420~\cite{impinj}) can read more than 1000+ tags per second, which is enough for the nanodrone swarm with hundreds of nanodrones. Moreover, the parallel decoding techniques described in the above subsections could further improve the sensing ability of RFID systems to decrease the processing delay.

\noindent \textbf{Long-Range Sensing.} \sysname's design is based on the commodity passive RFID system, whose operation range is limited by the reader's ability to activate the battery-free RFID tags. The commodity passive RFID system has an operation range of 5-15m at best~\cite{wang2019pushing} due to the low efficiency of energy harvesting. However, the short communication range is not inherited from \sysname's design, and we can always use the battery-instrumented tags (e.g., solar-powered tag) or tunneling RFID tag~\cite{amato2018tunneling, varshney2020tunnel} for long-range sensing based on our architecture, design and 3D nanodrone swarm pattern recognition algorithm. Moreover, recent research~\cite{wang2019pushing} shows that we can increase battery-free RFID tags' communication range by deploying multiple readers.

\section{Conclusion}
\label{sec:cons}

In this paper, we propose \sysname, a system that can predict the relative position of the nanodrones in the swarm and further recognize the 3D swarm pattern, using the commodity passive RFID tags. To do so, we attach a commodity passive RFID tag on each nanodrone, such that the relative position of the nanodrones in the swarm can be estimated by profiling the spatial-temporal phase readings using machine learning. We believe \sysname can proliferate the human-drone interaction, dazzling drone shows for the entertainment industry, and safe nanodrone swarm platooning.  

\bibliographystyle{IEEEtran}

\bibliography{references}

\end{document}